\input harvmac.tex
\input epsf
\input graphicx

\noblackbox
\def\figin{\epsfcheck\figin}\def\figins{\epsfcheck\figins}
\def\epsfcheck{\ifx\epsfbox\UnDeFiNeD
\message{(NO epsf.tex, FIGURES WILL BE IGNORED)}
\gdef\figin##1{\vskip2in}\gdef\figins##1{\hskip.5in}
\else\message{(FIGURES WILL BE INCLUDED)}%
\gdef\figin##1{##1}\gdef\figins##1{##1}\fi}
\def\DefWarn#1{}
\def\figinsert{\goodbreak\midinsert}
\def\ifig#1#2#3{\DefWarn#1\xdef#1{fig.~\the\figno}
\writedef{#1\leftbracket fig.\noexpand~\the\figno}%
\figinsert\figin{\centerline{#3}}\medskip\centerline{\vbox{\baselineskip12pt
\advance\hsize by -1truein\noindent\footnotefont{\bf Fig.~\the\figno } \it#2}}
\bigskip\endinsert\global\advance\figno by1}


\def\encadremath#1{\vbox{\hrule\hbox{\vrule\kern8pt\vbox{\kern8pt
 \hbox{$\displaystyle #1$}\kern8pt}
 \kern8pt\vrule}\hrule}}
 %
 %
 

 \font\cmss=cmss10
 \font\cmsss=cmss10 at 7pt
 \def\rlx{\relax\leavevmode}
 \def\inbar{\vrule height1.5ex width.4pt depth0pt}
 \def\IC{\relax\,\hbox{$\inbar\kern-.3em{\rm C}$}}
 \def\IN{\relax{\rm I\kern-.18em N}}
 \def\IP{\relax{\rm I\kern-.18em P}}

\def\ZZ{\rlx\leavevmode\ifmmode\mathchoice{\hbox{\cmss Z\kern-.4em Z}}
  {\hbox{\cmss Z\kern-.4em Z}}{\lower.9pt\hbox{\cmsss Z\kern-.36em Z}}
  {\lower1.2pt\hbox{\cmsss Z\kern-.36em Z}}\else{\cmss Z\kern-.4em Z}\fi}
 \def\IZ{\relax\ifmmode\mathchoice
 {\hbox{\cmss Z\kern-.4em Z}}{\hbox{\cmss Z\kern-.4em Z}}
 {\lower.9pt\hbox{\cmsss Z\kern-.4em Z}}
 {\lower1.2pt\hbox{\cmsss Z\kern-.4em Z}}\else{\cmss Z\kern-.4em Z}\fi}
 \def\IZ{\relax\ifmmode\mathchoice
 {\hbox{\cmss Z\kern-.4em Z}}{\hbox{\cmss Z\kern-.4em Z}}
 {\lower.9pt\hbox{\cmsss Z\kern-.4em Z}}
 {\lower1.2pt\hbox{\cmsss Z\kern-.4em Z}}\else{\cmss Z\kern-.4em Z}\fi}

 \def\narrowplus{\kern -.04truein + \kern -.03truein}
 \def\narrowminus{- \kern -.04truein}
 \def\narrowminussub{\kern -.02truein - \kern -.01truein}

 \def\ep{{\epsilon}}

 \def\frac#1#2{{#1\over #2}}

 \def\IZ{\relax\ifmmode\mathchoice
 {\hbox{\cmss Z\kern-.4em Z}}{\hbox{\cmss Z\kern-.4em Z}}
 {\lower.9pt\hbox{\cmsss Z\kern-.4em Z}}
 {\lower1.2pt\hbox{\cmsss Z\kern-.4em Z}}\else{\cmss Z\kern-.4em Z}\fi}
 \def\IB{\relax{\rm I\kern-.18em B}}
 \def\IC{{\relax\hbox{$\inbar\kern-.3em{\rm C}$}}}
 \def\Ic{{\relax\hbox{$\inbar\kern-.22em{\rm c}$}}}
 \def\ID{\relax{\rm I\kern-.18em D}}
 \def\IE{\relax{\rm I\kern-.18em E}}
 \def\IF{\relax{\rm I\kern-.18em F}}
 \def\IG{\relax\hbox{$\inbar\kern-.3em{\rm G}$}}
 \def\IGa{\relax\hbox{${\rm I}\kern-.18em\Gamma$}}
 \def\IH{\relax{\rm I\kern-.18em H}}
 \def\II{\relax{\rm I\kern-.18em I}}
 \def\IK{\relax{\rm I\kern-.18em K}}
 \def\IP{\relax{\rm I\kern-.18em P}}

 \font\cmss=cmss10 \font\cmsss=cmss10 at 7pt
 \def\IR{\relax{\rm I\kern-.18em R}}

 %

 %
 %
 \def\eqnn#1{\xdef
#1{(\secsym\the\meqno)}\writedef{#1\leftbracket#1}%
 \global\advance\meqno by1\wrlabeL#1}
 \def\eqna#1{\xdef
#1##1{\hbox{$(\secsym\the\meqno##1)$}}

\writedef{#1\numbersign1\leftbracket#1{\numbersign1}}%
 \global\advance\meqno by1\wrlabeL{#1$\{\}$}}
 \def\eqn#1#2{\xdef
#1{(\secsym\the\meqno)}\writedef{#1\leftbracket#1}%
 \global\advance\meqno by1$$#2\eqno#1\eqlabeL#1$$}

\newdimen\tableauside\tableauside=1.0ex
\newdimen\tableaurule\tableaurule=0.4pt
\newdimen\tableaustep
\def\phantomhrule#1{\hbox{\vbox to0pt{\hrule height\tableaurule width#1\vss}}}
\def\phantomvrule#1{\vbox{\hbox to0pt{\vrule width\tableaurule height#1\hss}}}
\def\sqr{\vbox{%
  \phantomhrule\tableaustep
  \hbox{\phantomvrule\tableaustep\kern\tableaustep\phantomvrule\tableaustep}%
  \hbox{\vbox{\phantomhrule\tableauside}\kern-\tableaurule}}}
\def\squares#1{\hbox{\count0=#1\noindent\loop\sqr
  \advance\count0 by-1 \ifnum\count0>0\repeat}}
\def\tableau#1{\vcenter{\offinterlineskip
  \tableaustep=\tableauside\advance\tableaustep by-\tableaurule
  \kern\normallineskip\hbox
    {\kern\normallineskip\vbox
      {\gettableau#1 0 }%
     \kern\normallineskip\kern\tableaurule}%
  \kern\normallineskip\kern\tableaurule}}
\def\gettableau#1 {\ifnum#1=0\let\next=\null\else
  \squares{#1}\let\next=\gettableau\fi\next}

\tableauside=1.0ex
\tableaurule=0.4pt

\def\IE{\relax{\rm I\kern-.18em E}}
\def\IP{\relax{\rm I\kern-.18em P}}

\lref\tv{
  M.~Aganagic, A.~Klemm, M.~Marino, C.~Vafa,
  ``The Topological vertex,''
Commun.\ Math.\ Phys.\  {\bf 254}, 425-478 (2005).
[hep-th/0305132].
}

\lref\dvI{
R.~Dijkgraaf and C.~Vafa,
``Matrix models, topological strings, and supersymmetric gauge theories,''
Nucl.\ Phys.\ B {\bf 644}, 3 (2002)
[arXiv:hep-th/0206255].
}

\lref\dvII{
R.~Dijkgraaf and C.~Vafa,
``On geometry and matrix models,''
Nucl.\ Phys.\ B {\bf 644}, 21 (2002)
[arXiv:hep-th/0207106].
}

\lref\akmv{
M.~Aganagic, A.~Klemm, M.~Mari\~no and C.~Vafa,
``The topological vertex,''
arXiv:hep-th/0305132.
}

\lref\akv{
  M.~Aganagic, A.~Klemm and C.~Vafa,
  ``Disk Instantons, Mirror Symmetry and the Duality Web,''
Z.Naturforsch. A57 (2002) 1-28, [hep-th/0105045].
}

\lref\GGV{
  L.~F.~Alday, D.~Gaiotto, S.~Gukov, Y.~Tachikawa, H.~Verlinde,
  ``Loop and surface operators in N=2 gauge theory and Liouville modular geometry,''
JHEP {\bf 1001}, 113 (2010).
[arXiv:0909.0945 [hep-th]].
}

\lref\adkmv{ M.~Aganagic, R.~Dijkgraaf, A.~Klemm, M.~Marino, C.~Vafa,
  ``Topological strings and integrable hierarchies,''
Commun.\ Math.\ Phys.\  {\bf 261}, 451-516 (2006).
[hep-th/0312085].
}

\lref\no{
N.~Nekrasov and A.~Okounkov,
``Seiberg-Witten theory and random partitions,''
arXiv:hep-th/0306238.
}

\lref\VerlindeSN{
  E.~P.~Verlinde,
  ``Fusion Rules and Modular Transformations in 2D Conformal Field Theory,''
Nucl.\ Phys.\  {\bf B300}, 360 (1988).
}\lref\ms{
  G.~W.~Moore, N.~Seiberg,
  ``Lectures On Rcft,''
}

\lref\toda{
  R.~Dijkgraaf, C.~Vafa,
  ``Toda Theories, Matrix Models, Topological Strings, and N=2 Gauge Systems,''
[arXiv:0909.2453 [hep-th]].
}

\lref\mmb{M. Marino, "Chern-Simons Theory, Matrix Models, And Topological Strings" (Oxford University Press, 2005).}

\lref\lmv{J. M. F. Labastida, M. Marino, and C. Vafa, ÒKnots, Links, and Branes At Large N,Ó JHEP 0011(2000) 007, hep-th/0010102.}

 \lref\AY{M. Aganagic and M. Yamazaki, ÒOpen BPS Wall Crossing and M-theory,Ó Nucl. Phys. B834 (2010) 258Ð272, 0911.5342.}

\lref\ems{
  S.~Elitzur, G.~W.~Moore, A.~Schwimmer, N.~Seiberg,
  ``Remarks on the Canonical Quantization of the Chern-Simons-Witten Theory,''
Nucl.\ Phys.\  {\bf B326}, 108 (1989).
}
\lref\mm{
  M.~Marino,
  ``Chern-Simons theory, matrix integrals, and perturbative three manifold invariants,''
Commun.\ Math.\ Phys.\  {\bf 253}, 25-49 (2004).
[hep-th/0207096].
}

\lref\ovknot{H.~Ooguri and C.~Vafa,
``Knot invariants and topological strings,''
Nucl.\ Phys.\ B {\bf 577}, 419 (2000)
[arXiv:hep-th/9912123].
}

\lref\gr{N. M. Dunfield, S. Gukov, and J. Rasmussen, ÒThe Superpotential For Knot Homologies,Ó Experiment. Math. 15 (2006) 129, math/0505662.}

\lref\rasb{J.Rasmussen, "Some differentials on Khovanov-Rozansky homology", arXiv:math/0607544.}

\lref\rasa{J.Rasmussen, "Khovanov-Rozansky homology of two-bridge knots and links",
arXiv:math.GT/0508510.}

\lref\gvI{R.~Gopakumar and C.~Vafa,
``On the gauge theory/geometry correspondence,''
Adv.\ Theor.\ Math.\ Phys.\  {\bf 3}, 1415 (1999)
[arXiv:hep-th/9811131].
}
\lref\dimofte{
  T.~Dimofte, S.~Gukov, L.~Hollands,
  ``Vortex Counting and Lagrangian 3-manifolds,''
[arXiv:1006.0977 [hep-th]].
}
\lref\taki{
  M.~Taki,
  ``Flop Invariance of Refined Topological Vertex and Link Homologies,''
[arXiv:0805.0336 [hep-th]].
}

\lref\gvII{
  R.~Gopakumar, C.~Vafa,
  ``M theory and topological strings. 1.,''
[hep-th/9809187];
  R.~Gopakumar, C.~Vafa,
  ``M theory and topological strings. 2.,''
[hep-th/9812127].
}

\lref\wrr{
D. Gaiotto and E. Witten, ÒSupersymmetric Boundary Conditions In N = 4 Super Yang-Mills Theory,Ó J. Stat. Phys. (2009) 135 789-855. arXiv:0804.2902.}

\lref\wrrr{ D. Gaiotto and E. Witten, ÒJanus Configurations, Chern-Simons Couplings, And The Theta-Angle in N=4 Super Yang-Mills Theory,Ó JHEP 1006 2010 097, arXiv:0804.2907.}

\lref\wp{E. Witten, ÒA New Look At The Path Integral Of Quantum Mechanics,Ó arXiv:1009.6032.}

\lref\nikitaa{
N.~A.~Nekrasov,
``Seiberg-Witten prepotential from instanton counting,''
arXiv:hep-th/0206161.
}
%


\lref\kh{
M. Khovanov, ÒA Categorification Of The Jones Polynomial,Ó Duke. Math. J. 101 (2000) 359-426.}

\lref\hiv{
T.~J.~Hollowood, A.~Iqbal, C.~Vafa,
``Matrix models, geometric engineering and elliptic genera,''
JHEP {\bf 0803}, 069 (2008).
[hep-th/0310272].
}

\lref\macda{I.G. Macdonald, "A new class of symmetric functions", Publ. I.R.M.A. Strasbourg, 372/S- 20, Actes 20 Seminaire Lotharingien (1988), 131-171.}

\lref\macdb{I.G. Macdonald, "Orthogonal polynomials associated with root systems", preprint (1988).}


\lref\ek{P. Etingof and A. Kirillov, Jr., "On Cherednik-Macdonald-Mehta identities", q-alg 9712051, Electr.Res.An., v.4(1998),p.43-47.
}

\lref\eka{P.I. Etingof and A.A. Kirillov, "Macdonald's polynomials and representations of quantum groups", Math. Res. Let. 1 (1994), 279Ð296}

\lref\ekb{P.I. Etingof and A.A. Kirillov, "Representation-theoretic proof of inner product and symmetry identities for Macdonald's polynomials", hep-th/9410169, to appear in Comp. Math. (1995).}

\lref\kirila{A. Kirillov, Jr., "Lectures on affine Hecke algebras and Macdonald's conjectures", Bull. Amer. Math. Soc. 34 (1997), 251Ð292.}

\lref\kirillov{ A. Kirillov, Jr.,
"On inner product in modular tensor categories. I",
 arXiv:q-alg/9508017}

\lref\cheredniko{I. Cherednik, "Macdonald's Evaluation Conjectures and Difference Fourier Transform", arXiv:q-alg/9412016}

\lref\cherednikt{ I. Cherednik, "Double Affine Hecke Algebras and Macdonald's Conjectures", The Annals of Mathematics,
Second Series, Vol. 141, No. 1 (Jan., 1995), pp. 191-216}

\lref\co{I. Cherednik and V. Ostrik, "From Double Affine Hecke Algebra to Fourier Transform", Selecta Math. (N.S.) 9, no. 2, 161249, (2003).}

\lref\taubes{C. Taubes, ÒLagrangians for the Gopakumar-Vafa conjecture,Ó math.DG/0201219.}

\lref\refmm{
  A.~Brini, M.~Marino, S.~Stevan,
  ``The Uses of the refined matrix model recursion,''
[arXiv:1010.1210 [hep-th]].
}

\lref\bcov{
M.~Bershadsky, S.~Cecotti, H.~Ooguri and C.~Vafa,
``Kodaira-Spencer theory of gravity and exact results for quantum string amplitudes,''
hep-th/9309140,
Commun.\ Math.\ Phys.\  {\bf 165} (1994) 311.}


\lref\amv{M.~Aganagic, M.~Mari\~no and C.~Vafa,
``All loop topological string amplitudes from Chern-Simons theory,''
hep-th/0206164.}

\lref\akmv{
  M.~Aganagic, A.~Klemm, M.~Marino, C.~Vafa,
  ``Matrix model as a mirror of Chern-Simons theory,''
JHEP {\bf 0402}, 010 (2004).
[hep-th/0211098].
}

\lref\gsv{
S. Gukov, A. S. Schwarz, and C. Vafa, ÒKhovanov-Rozansky Homology And Topological Strings,Ó Lett. Math. Phys. 74 (2005) 53-74, hep-th/0412243.
}

\lref\DVV{
  R.~Dijkgraaf, C.~Vafa, E.~Verlinde,
  ``M-theory and a topological string duality,''
[hep-th/0602087].
}

\lref\macdonald{I.G. Macdonald, {\it Symmetric functions and Hall polynomials},
Oxford University Press, 1995.}

  \lref\dbn{D. Bar-Natan, On Khovanov's Categorification Of The Jones Polynomial, arXiv:math/0201043, Alg. Geom. Topology 2 (2002) 337-370.}

  \lref\tur{P. Turner, "Five Lectures on Khovanov Homology"
 arXiv:math/0606464.}

 \lref\ak{M. Asaeda and M. Khovanov,  "Notes on link homology", arXiv/0804.1279.}

\lref\hk{R.S. Huerfano and M. Khovanov,
 "Categorification of some level two representations of sl(n)", arXiv:math/0204333.} \lref\kc{M. Khovanov,
 "Categorifications of the colored Jones polynomial",
arXiv:math/0302060.}\lref\web{B. Webster,
 "Knot invariants and higher representation theory I: diagrammatic and geometric categorification of tensor products", arXiv/1001.2020.}

 \lref\ps{P. Seidel and I. Smith, ÒA Link Invariant From The Symplectic Geometry Of Nilpotent Slices,Ó arXiv:math/0405089.}

 \lref\km{P. B. Kronheimer and T. S. Mrowka, ÒKnot Homology Groups From Instantons,Ó arXiv:0806.1053}

 \lref\kmt{ P. B. Kronheimer and T. S. Mrowka, ÒKhovanov Homology Is An Unknot-Detector,Ó arXiv:1005.4346}

\lref\ik{
A.~Iqbal and A.~K.~Kashani-Poor,
``$SU(N)$ geometries and topological string amplitudes,''
arXiv:hep-th/0306032.
}

\lref\jones{
V.F.R. Jones "Index for subfactors", Invent. Math. 72 (1983) 1-25;  V.F.R. Jones "A polynomial invariant for knots via von Neumann algebras", Bull. Amer. Math. Soc. 12 (1985) 103-112; V.F.R. Jones, "Hecke algebra representations of braid groups and link polynomials". Ann. of Math. (2) 126 (1987), no. 2, 335-388.}

\lref\wcs{
  E.~Witten,
  ``Quantum Field Theory and the Jones Polynomial,''
Commun.\ Math.\ Phys.\  {\bf 121}, 351 (1989).
}

\lref\giv{S. Gukov, A. Iqbal, C. Kozcaz, and C. Vafa, ÒLink Homologies and the Refined Topo- logical Vertex,Ó arXiv:0705.1368.}

\lref\civ{A. Iqbal, C. Kozcaz, C. Vafa, ÓThe Refined Topological VertexÓ, hep-th/0701156.}

\lref\wst{
  E.~Witten,
 ``Chern-Simons gauge theory as a string theory,''
Prog.\ Math.\  {\bf 133}, 637-678 (1995).
[hep-th/9207094].
}

\lref\wr{
  E.~Witten,
  ``Fivebranes and Knots,''
[arXiv:1101.3216 [hep-th]].
}

\lref\wrg{ D. Gaiotto and E. Witten, ``Knot Invariants from Four-Dimensional Gauge Theory,'' arXiv:1106.4789}

\lref\wrw{ E. Witten, ``Khovanov Homology And Gauge Theory,'' arXiv:1108.3103}

\lref\acdkv{
  M.~Aganagic, M.~C.~N.~Cheng, R.~Dijkgraaf, D.~Krefl, C.~Vafa,
  ``Quantum Geometry of Refined Topological Strings,''
[arXiv:1105.0630 [hep-th]].
}
\def\Title#1#2{\nopagenumbers\abstractfont\hsize=\hstitle\rightline{#1}
\vskip .5in\centerline{\titlefont #2}\abstractfont\vskip .5in\pageno=0}

{
\Title
{\vbox{
 \baselineskip12pt
}}
{\vbox{
\centerline{Refined Chern-Simons Theory}
\vskip 0.4 cm
\centerline{and }
\vskip 0.4cm
\centerline{Knot Homology }
}}
 \centerline{{\bf Mina Aganagic}$^{a,b}$ and {\bf Shamil Shakirov}$^{a,c}$}
\vskip 0.5cm

\centerline{$^a$ \it Department of Mathematics, University of California, Berkeley, USA}
\centerline{$^b$ \it Center for Theoretical Physics, University of California, Berkeley, USA}
\centerline{$^c$ \it Institute for Theoretical and Experimental Physics, Moscow, Russia}

\smallskip
\vskip 0.1cm
\centerline{\bf Abstract}
\vskip 0.2cm

The refined Chern-Simons theory is a one-parameter deformation of the ordinary Chern-Simons theory on Seifert manifolds. It is defined via an index of the theory on $N$ M5 branes, where the corresponding one-parameter deformation is a natural deformation of the geometric background. Analogously with the unrefined case, the solution of refined Chern-Simons theory is given in terms of $S$ and $T$ matrices, which are the proper Macdonald deformations of the usual ones. This provides a direct way to compute refined Chern-Simons invariants of a wide class of three-manifolds and knots. The knot invariants of refined Chern-Simons theory are conjectured to coincide with the knot superpolynomials -- Poincare polynomials of the triply graded knot homology theory. This conjecture is checked for a large number of torus knots in $S^3$, colored by the fundamental representation. This is a short, expository version of arXiv:1105.5117, with some new results included.\foot{Based on talks presented by M.A. at several conferences and workshops, including String-Math 2011 Conference at U. of Pennsilvania.}


\goodbreak
\vfill
\eject
}
\newsec{Introduction}

One of the beautiful stories in the marriage of mathematics and physics developed from Witten's realization \wcs\ that
three dimensional Chern-Simons theory on $S^3$ computes the polynomial invariant of knots constructed by Jones in \jones . While Jones constructed an invariant $J(K, {\bf q})$ of knots in three dimensions, his construction relied on projections of knots to two dimensions. This obscured the three dimensional origin of the Jones polynomial.
The fact that Chern-Simons theory is a topological quantum field theory in three dimensions made it  manifest that the Jones polynomial is an invariant of the knot, and independent of the two dimensional projection. Moreover, it also gave rise to new topological invariants of three manifolds and knots in them. For any three manifold $M$ and a knot in it, Chern Simons path integral, with Wilson loop observable inserted along the knot, gives a topological invariant that depends only on $M$, $K$ and the representation of the gauge group. Moreover, Chern-Simons theory gives a whole family of invariants associated to $M$ and $K$, by changing the gauge group $G$ and the representation $R$ on the Wilson line. Jones polynomial $J(K, {\bf q})$ corresponds to $G=SU(2)$, and $R$ the fundamental, two dimensional representation of $G$.  Taking $G=SU(n)$ instead, one computes the HOMFLY polynomial $H(K, {\bf q}, {\bf a})$ \ref\HOMFLY{P. Freyd, D. Yetter, J. Hoste, W.B.R. Lickorish, K. Millett, and A. Ocneanu, "A New Polynomial Invariant of Knots and Links". Bulletin of the American Mathematical Society 12 (2) (1985) 239-246. doi:10.1090/S0273-0979-1985-15361-3.} evaluated at ${\bf a}={\bf q}^n.$
The work in \wcs\ was made even more remarkable by the fact that it explained how to solve Chern-Simons theory for any $M$ and collection of knots in it.

A mystery left open by \wcs\ is the integrality of the coefficients of the Jones and HOMFLY polynomials. They are both Laurent polynomials in ${\bf q}$, and in the latter case ${\bf a}$, with integer coefficients. While Chern-Simons theory gives means of computing knot invariants, it gives no insight into question why the coefficients are integers. An answer to this question was provided by \kh .  Khovanov associates a bi-graded (co)homology theory to a knot $H^{i,j}(K)$, in such a way that its Euler characteristic is the Jones polynomial,
$$J(K,{\bf q}) = \sum_{i,j} (-1)^i{\bf q}^j\, {\rm dim}\,H^{i,j}(K).$$
Interpreted in this way, the integrality of the coefficients is manifest, since they are counting dimensions of knot homology groups. This gives rise to a refinement of the Jones  polynomial, where one computes the Poincare polynomial instead,
$$
Kh(K, {\bf q}, {\bf t}) =    \sum_{i,j} {\bf t}^i{\bf q}^j\, {\rm dim}\,H^{i,j}(K).
$$
This depends on one extra parameter ${\bf t},$ and reduces to the Jones polynomial at ${\bf t}=-1$. Later, many generalizations of \kh\ were constructed. In particular generalization to $SU(N)$ knot invariants was constructed by Khovanov and Rozansky
\ref\kro{M. Khovanov and L. Rozansky, "Matrix factorizations and link homology I", arXiv:math/0401268.}
\ref\krt{M. Khovanov and L. Rozansky, "Matrix factorizations and link homology II", arXiv:math/0505056.}. Knot homology theory corresponding to refinement of the HOMFLY polynomial was conjectured to exist by \gr. The resulting three variable Poincare polynomial was named the superpolynomial  \gr.

Chern-Simons theory arises naturally in the context of string theory, \wst.  Namely, the $SU(N)$ Chern-Simons partition function on $M$ is the same as the partition function of $N$ M5 branes wrapping $M\times {\bf C}\times S^1$ in M-theory on $(T^*M\times \,{\IC}^2\times S^1)_q$ where, as one goes around the $S^1$, the two ${\,\IC}$ planes get rotated by
$$(z_1, z_2) \rightarrow ( q z_1,  q^{-1}z_2).
$$
The M-theory partition function is an index, ${\rm Tr}(-1)^F q^{S_1-S_2}$, where $S_1$ and $S_2$ are the generators of rotations around the two complex planes. As explained in \gsv\ it is expected that knot homologies arise from string theory by counting BPS states in M-theory where one keeps track of both quantum numbers $S_1$ and $S_2$ separately.

In \ref\AS{
  M.~Aganagic and S.~Shakirov,
  ``Knot Homology from Refined Chern-Simons Theory,''
[arXiv:1105.5117 [hep-th]].
},  we argued that, provided that the three manifold $M$ and knots in it possess enough symmetry (this is the case when $M$ is a Seifert manifold with Seifert knots), one can construct a refined index that keeps track of the spins $S_1$ and $S_2$ separately. This corresponds to M-theory partition function in the background where $z_1$ and $z_2$ are allowed to rotate independently,
\eqn\rotate{
(z_1, z_2) \rightarrow  ( q z_1,  t^{-1} z_2).
}
Here $q=e^{\epsilon_1}, t = e^{-\epsilon_2}$, in terms of the usual parameters $\epsilon_{1,2}$ defining Nekrasov background \nikitaa , and the refined topological string.
The M-theory partition function is ${\rm Tr}(-1)^F q^{S_1-S_r} t^{S_r-S_R}$, where $S_r$ is the generator of the extra $U(1)$ symmetry present. Because of the additional grade, and since one is computing an index, the knot invariants of refined Chern-Simons theory need not be the same as the Poincare polynomials of knot homology. However, as we will see in some cases, the BPS states have $S_r=0$, there are no cancellations, and the two are equal.

This short note is devoted to overview of the results of \AS, and its organization is as follows. In section 2 we review the relation of the ordinary and the refined Chern-Simons theory. We explain how to compute the knot and the three manifold invariants of the refined Chern-Simons theory. We discuss the conjecture relating refined Chern-Simons knot invariants to the superpolynomial of \gr.
In section 3, we explain the M-theory definition of the refined Chern-Simons theory. We also discuss relation of the present constructions to \wr . In section 4 we explain how to compute the $S$ and the $T$ matrices, from M-theory.
In section 5 we show that the large $N$ dual of the refined Chern-Simons theory on the $S^3$ is the refined topological string on $X= {\cal O}(-1)\oplus {\cal O}(-1) \rightarrow {\IP}^1$.  In section 6, we finish with comments on relations to previous work and interesting directions of generalization. We add an appendix, with the explicit refined Chern-Simons knot invariants for some more complicated torus knots.

\newsec{Refined and Ordinary Chern-Simons Theory}

In \AS\ we formulated a refinement of $SU(N)$ Chern-Simons theory, which we define on Seifert three manifolds, with Seifert knots. Seifert three manifolds are circle fibrations over a Riemann surface. They admit a (semi-)free $U(1)$ action, corresponding to rotating the $S^1$ fiber; the action of the $U(1)$ is free except that a discrete subgroup of $U(1)$ can act with fixed points. Seifert knots are knots wrapping the $S^1$ fiber, and projecting to points on the Riemann surface.

Refined Chern-Simons theory is a topological theory in three dimensions. In any such theory, all amplitudes, corresponding to any three manifold, with arbitrary knots, can be written in terms of three building blocks: the $S$, $T$ and the braiding matrix $B$. In refined Chern-Simons theory, only a subset of amplitudes enter, as only those preserve the $U(1)$ symmetry. Moreover, the $S$ and the $T$ matrices of the refined Chern-Simons theory are a one parameter deformation of those in ordinary Chern-Simons theory. The $S$ and the $T$ matrices provide a unitary representation
of the modular $SL(2, {\bf Z})$ group on the Hilbert space ${\cal H}_{T^2}$ of the theory on a torus: they satisfy
\eqn\sltz{ S^4=1, \qquad \qquad (ST)^3=S^2.
}
A basis of the Hilbert space ${\cal H}_{T^2}$ can be obtained by taking a solid torus and placing Wilson lines in various representations in its interior in a particular way. More precisely, choosing a basis of $H_1(T^2)$ of the boundary torus, and taking the $(1,0)$ cycle of the $T^2$ to be contractible in the interior, one defines a state
\eqn\hilb{
|R_i \rangle \;\; \in \;\; {\cal H}_{T^2}
}
by the path integral on the solid torus with a  Wilson line in representation $R_i$ running along the $(0,1)$ cycle of the torus.
Moreover, on the boundary of the $T^2$, one gets the action of $SL(2,\IZ)$ corresponding to the mapping class group of the torus. An element
$K$  of $SL(2, \IZ)$ acts on the basis states by
$$
K |R_i\rangle  = \sum_j {K^j}_i |R_j\rangle
$$
simply corresponding to the fact that the Hilbert space is finite dimensional, and $K$ acts on it.
The representation of $SL(2,\IZ)$ acting on ${\cal H}_{T^2}$, is generated by $S$ and $T$ matrices,
$$
{S^{i}}_j, \qquad {T^{i}}_{j},
$$
satisfying the defining relations of $SL(2,\IZ)$,
\eqn\sltz{
S^4= 1,\qquad (ST)^{3}= S^2.
}
 Let
$$
K_{{\bar i} j}= \langle R_j | K R_i \rangle.
$$
This has has to be unitary,
\eqn\unit{
K_{{\bar i} j}^* = {K^{-1}}_{{\bar j} i},
}
since otherwise
3d general covariance would have been lost.\foot{This follows from $\langle K R_j | R_i \rangle = \langle R_j | K^{-1}R_i \rangle$ using
$\langle K R_j | R_i \rangle=\langle R_i| K R_j  \rangle^*$.}
Topological invariance further constrains the representation. For example, the $S$ matrix has to be symmetric and satisfy,
\eqn\ssym{
S^{-1}_{{\bar i}j} = S^*_{{\bar i} j}.
}
This follows since $S_{{\bar i}j}$ is the amplitude of a Hopf link in $S^3$, obtained by gluing two solid tori, with
Wilson lines corresponding to states $|R_i\rangle$ and $|R_j\rangle $ we defined before. Gluing these with
an $S$ transformation of the boundary we get the amplitude corresponding to two linked knots in the $S^3,$ the Hopf link:
$$S_{{\bar i} j} = \langle R_i | S | R_j\rangle.
$$
The fact that we can smoothly re-arrange the link so that the roles of $R_i$ and $R_j$ get exchanged implies that $S$ has to be symmetric, $S_{{\bar i} j} = S_{{\bar j} i}$. Unitarity then implies
\ssym.

Indices are raised and lowered by a hermitian metric $g$, defined by
\eqn\met{
g_{{\bar i} j} = \langle R_i | R_j \rangle=Z(S^2 \times S^1, {\bar R_i} , R_j).
}
This corresponds to taking two solid tori, with Wilson lines in representations ${R_i}$, $R_j$ inside, and gluing together, with trivial identifications. The three manifold one obtains is $S^2\times S^1$, with two Wilson lines.
The metric is hermitian, since exchanging the roles of $R_i$ and $R_j$ corresponds to orientation reversal of the manifold,
$$
g_{{\bar j} i}= \langle R_j | R_i \rangle = \langle R_i | R_j \rangle^* = g_{{\bar i} j}^*,
$$
where $^*$ denotes complex conjugation.
We have, using the definitions,
$$
K_{{\bar j} i} =  \sum_k {K^k}_i \,g_{{\bar j} k}.
$$
%

%
%

\subsec{"Ordinary" Chern-Simons Theory}

In the case of $SU(N)$ Chern-Simons theory, the basis of Hilbert space ${\cal H}_{T^2}$ is provided by the conformal blocks on $T^2$ of $SU(N)_k$ $WZW$ model \wcs .  Only a subset of $SU(N)$ representations enter, those that correspond to integrable highest weight representations of the affine lie algebra. This can be phrased in terms of a constraint on the corresponding Young diagram that $0\leq R_1 \leq k$, where $R_1$ is the length of the first row.
%
%
The basis of the Hilbert space \hilb\ provided by the path integral on the solid torus with Wilson lines, is automatically orthonormal \wcs :
\eqn\ort{
g_{{\bar i}j } = \langle R_i|R_j\rangle = {\delta^i}_j.
}
The $S$ and $T$ matrices are given by
$$
S_{{\bar i}j}/S_{00} =s_{R_i}(q^{\rho})s_{R_j}(q^{\rho +{R_i}}),
$$
where $s_{R_i}(x)$ are Schur functions, and
$$T^{i}_{ j} = {\delta^i}_j q^{{1\over 2}({R_i} + 2 \rho, {R_i})-{k\over 2 N}(\rho,\rho)},
$$
with
$$
S_{00}= {{i}^{N(N-1)/2}\over N^{1\over 2}(k+N)^{N-1\over 2}}
\; \prod_{\alpha>0}(q^{-(\alpha, \rho)/2} - q^{(\alpha, \rho)/2}) \
$$
where the product is over all positive roots $\alpha$, and ${R_i}$ denotes the highest weight of the corresponding representation $R_i$.
When $q$ is a root of unity, determined by the level $k$ and the rank $N$ of the Chern-Simons gauge group,
$$
q = e^{2\pi i \over k+N},
$$
the matrices $S$ and $T$ provide a unitary representation of $SL(2, {\bf Z})$, in particular, satisfy the properties \sltz\unit\ssym .

\subsec{Refined Chern-Simons Theory}

The refined Chern-Simons  theory depends on an extra parameter, but shares many of the same properties as the ordinary Chern-Simons theory. To specify the theory we need to specify the rank $N$ and the level $k$, $N,k\in Z_{>0}$, together with one additional parameter $\beta \in {\bf R}$. The theory at level $k$ has the Hilbert space ${\cal H}_T^2$ of the same dimension as ordinary Chern-Simons theory, with a basis labeled by integrable representations of $SU(N)_k$. The hermitian metric is diagonal\foot{We could have chosen a different normalization of $|R_i\rangle$, that would have made the basis orthonormal. We chose to stick to the conventions of \AS\ and not do that.}
$$\langle R_i |R_j\rangle =g_{{\bar i} j} = g_i {\delta^{i}}_j.
$$
Set
$$t= q^{\beta}.
$$
While the theory makes sense for any $\beta$, the expressions are the simplest for $\beta\in {\bf Z}_{>0}$. We then get
\eqn\metd{
g_i = \prod_{m=0}^{\beta -1}
\prod_{\alpha>0}{q^{-{1\over 2}( \alpha, R_i)}t^{ -{1\over 2} (\alpha,\rho)} q^{-{m\over 2}}-
q^{{1\over 2} (\alpha, R_i)} t^{ {1\over 2} (\alpha,\rho)} q^{m\over 2}\over
q^{-{1\over 2}( \alpha, R_i)}t^{ -{1\over 2} (\alpha,\rho)} q^{m\over 2}-
q^{{1\over 2} (\alpha, R_i)} t^{ {1\over 2} (\alpha,\rho)} q^{-{m\over 2}}}
}
Setting
\eqn\rtu{
q = e^{2\pi i \over k+\beta N}, \qquad t = e^{2 \pi i \beta \over k+\beta N}.
}
the metric vanishes for representations other than the those whose Young tableau fits in box of width $k$, corresponding to representations of $SU(N)$ at level $k$. This is to be expected, since $\beta$ is arbitrary and we can change it away from the unrefined value $\beta=1$ adiabatically -- in a finite dimensional Hilbert space, the states have nowhere to go to.

The $S$ and the $T$ matrices are given by
\eqn\nsm{
S_{{\bar i}j}/S_{00} =M_{R_i}(t^{\rho})M_{R_j}(t^{\rho} q^{ {R_i}}),
}
where $M_{R_i}(x)$ are Macdonald polynomials, and
\eqn\ntm{T^{i}_ {j} = {\delta^i}_j \;q^{{1\over 2}({R_i}, {R_i})}\,t^{({R_i}, \rho)}t^{{\beta-1\over 2}(\rho, \rho)}q^{-{k\over 2 N}(\rho,\rho)},
}
where
$$
S_{00}= {{i}^{N(N-1)/2}\over N^{1\over 2}(k+\beta N)^{N-1\over 2}}
\; \prod_{m=0}^{\beta-1}\prod_{\alpha>0}(q^{-m/2} t^{-(\alpha,  \rho)/2} - q^{m/2} t^{(\alpha,  \rho)/2}) \;$$
Setting $\beta$ to $1$, $q$ and $t$ coincide, Macdonald polynomials become Schur functions, and the refined Chern-Simons amplitudes reduce to ordinary ones, as they should.

It is crucial, for the three dimensional interpretation of $S_{{\bar i} j}$ as the expectation value of the colored Hopf link, that it is symmetric. This fact is not obvious from the formula \nsm , just like in the unrefined case. The symmetry of $S_{{\bar i} j}$ is in fact
the content of one of the famous Macdonald conjectures \macda\macdb.  We will see that the M-theory derivation of $ S_{{\bar i} j}$ will result in an expression for $S$ that makes this symmetry manifest. Since the metric $g_{{\bar i} { j}}$ is not identity, $S_{{\bar i} j}$ and ${S^{i}}_j$ are not the same, instead,
$
S_{{\bar i} {j}} = \sum_{k} g_{{\bar i}k} {S^{k}}_j = g_{i} {S^{i}}_j,
$
and similarly for $T$.
In particular, while $S_{{\bar i}j}$ is symmetric, ${S^{i}}_{j}$, is not.
The fact that $S$ and $T$ to provide a unitary representation on $SL(2,\IZ)$,  was proven in \refs{\cheredniko, \kirillov, \cherednikt }  (for integer $\beta$). The fact that this holds for arbitrary beta can be checked in examples.

\subsec{Verlinde Coefficients}

One can define Verlinde coefficients $N_{ijk}$ by the partition function of the theory on $S^2\times S^1$ with Wilson lines in representations $R_i$, $R_j$ and $R_k$ inserted at three points on the $S^2$, and winding around the $S^1$.
\eqn\vcd{
N_{ijk} = Z(S^2\times S^1, R_i,R_j, R_k)=\langle 0 | R_i R_j R_k \rangle.
}
It can be shown, by computing the same amplitude in two different ways, in the refined and unrefined case alike, that $N_{ijk}$ satisfy the Verlinde formula \VerlindeSN\
\eqn\vf{
S_{{\bar k }i}\, S_{{\bar k} j}/S_{{\bar k} 0 } = \sum_{\ell}{ N^{\ell}}_{i j} S_{{\bar k }\ell}
}
or equivalently,
\eqn\vftwo{
N_{{i} { j} {\bar k}} = \sum_{\ell} {S_{{\bar\ell}i} \, S_{{\bar \ell}j}\,{{( S^*)_{\bar k}}^{{\bar \ell}}\,}/S_{{\bar \ell}0} }.
}
The formula \vf\ can be proven by noting \co\ that on the one hand, insertions of Wilson-loops are realized by multiplication by Schur/Macdonald polynomials in $N$ holonomy variables; and on the other hand,  the $S$ matrix is obtained by evaluating the Schur/Macdonald polynomials at special points.

\subsec{Three Manifold Invariants from Refined Chern-Simons Theory}

From $S$ and $T$, by cutting and gluing, we can obtain invariants of knots and three manifolds which preserve the $U(1)$ action. In particular, it is interesting to consider Seifert three manifolds without knots. A compact Seifert three manifold, fibered over the Riemann surface of genus $g$, carries labels (the description of the geometry is borrowed from \ref\bw{
  C.~Beasley, E.~Witten,
  ``Non-Abelian localization for Chern-Simons theory,''
J.\ Diff.\ Geom.\  {\bf 70}, 183-323 (2005).
[hep-th/0503126].
})
$$
( g, n; (\alpha_1, \beta_1), \ldots,(\alpha_r,\beta_r)),
$$
where $n$ is the degree of the circle bundle, and $(\alpha_i,\beta_i)$ are integers, parameterizing the type of special fibers that occur.
%
%
The three manifold invariant of this can be computed\foot{See for example, \ref\hn{S. K. Hansen, "Reshetikhin-Turaev Invariants of Seifert 3-Manifolds and a Rational
Surgery Formula," Algebr. Geom. Topol. 1 (2001) 627Ð686, math.GT/0111057.} for details. Earlier work on three manifold invariants of Seifert spaces includes \lref\seifa{R. Lawrence and L. Rozansky, "Witten-Reshetikhin-Turaev Invariants of Seifert Manifolds," Commun. Math. Phys. 205 (1999) 287Ð314}
\lref\seifb{L. Rozansky, "Residue Formulas for the Large k Asymptotics of Witten's Invariants of Seifert Manifolds: The Case of SU(2)," Commun. Math. Phys. 178 (1996) 27Ð60, hep-th/9412075}
\lref\seifc{D. Freed and R. Gompf, "Computer Calculation of WittenÕs 3-Manifold Invariant," Commun. Math. Phys. 141 (1991) 79Ð117}
\lref\seifd{L. Jeffrey, "On Some Aspects of Chern-Simons Gauge Theory," D.Phil. thesis, University of Oxford, 1991}
\lref\seife{L. Jeffrey, "Chern-Simons-Witten Invariants of Lens Spaces and Torus Bundles," and the Semiclassical Approximation,  Commun. Math. Phys. 147 (1992) 563Ð604.}
\lref\seiff{ S. Garoufalidis, "Relations Among 3-Manifold Invariants," Ph.D. thesis, University of
Chicago, 1992.}
\lref\seifg{ J.R. Neil, "Combinatorial Calculation of the Various Normalizations of the Witten
Invariants for 3-Manifolds," J. Knot Theory Ramifications 1 (1992) 407Ð449.}
\lref\seifh{ L. Rozansky, "A Large k Asymptotics of Witten's Invariant of Seifert Manifolds,"
Comm. Math. Phys. 171 (1995) 279Ð322, hep-th/9303099.} \refs{\seifa,\seifb,\seifc,\seifd,\seife,\seiff,\seifg,\seifh}. }
$$
Z(g, n; (\alpha_1, \beta_1), \ldots,(\alpha_r,\beta_r)) = \sum_{j}T_j^{n}  (g_j)^{g-1}  (S_{0,j})^{2-r-2g}\Bigl( \prod_{i=1}^r (SK^{(\alpha_i, \beta_i)})_{0, j}\Bigr)
$$
where $K^{(\alpha_i, \beta_i)}$ is an $SL(2,{\IZ})$ matrix whose first column is ${(\alpha_i, \beta_i)}$. Each such matrix, as an element of $SL(2, \IZ)$ can be written as a product of $S$ and $T$ matrices. The $U(1)$ bundle over $\Sigma_g$ has first Chern class $n - \sum_{i=1}^r \beta_i/\alpha_i$. In the presence of special fibers, $\Sigma_g$ is an orbifold.

Written in terms of $S$ and $T$, the three manifold invariant no longer depends on the underlying topological field theory, but only on $M$. The dependence on the theory enters only through $S$ and $T$. To get a three manifold invariant corresponding to $SU(N)$ Chern-Simons theory on $M$, one would use the $S$ and the $T$ matrices of the $SU(N)_k$ WZW model. To get the path integral of the refined Chern-Simons theory on $M$ instead, one uses the refined $S$ and $T$ matrices of the previous subsection.

In the simple case without special fibers $(g,n)$, the three manifold is a circle bundle over a smooth Riemann surface $\Sigma_g$ of degree $n$, the corresponding partition function is
$$ Z(g,n) = \sum_j {  T_j^n(g_j)^{g-1}  (S_{0j})^{2-2g}}.
$$
We can obtain $(g, n; (\alpha_1, \beta_1), \ldots,(\alpha_r,\beta_r))$ from this by cutting out the neighborhoods of $r$ knots wrapping the $S^1$ fibers over points on $\Sigma$, and gluing back corresponding solid tori by $SL(2, {\IZ})$ transformation of their boundaries, corresponding to $K^{(\alpha_i, \beta_i)}$.

\subsec{Knot Invariants in Operator Formalism and the Refined Chern-Simons Theory}

The case of most interest for knot theory is $M=S^3$. $S^3$ can be viewed as a Seifert manifold in several different ways. The knots wrapping the Seifert fiber torus knots in $S^3$. We view the $S^3$ as a locus in ${\;\IC}^2$, with coordinates $z_1,z_2$, where
\eqn\thre{
|z_1|^2+|z_2|^2 = 1
}
An $(n,m)$ torus knot $K_{n,m}$ is described by \thre\ together with the equation
$$
z_1^n=z_2^m
$$
This is invariant under the $U(1)$ action that takes $(z_1, z_2)\rightarrow (\zeta^m z_1, \zeta^n z_2)$, with $\zeta=e^{i \theta}.$ This $U(1)$ action acts freely on the $S^3$, except for a ${\IZ}_m$ subgroup, generated by $\zeta= e^{2 \pi i /m}$, that has fixed points at $z_2=0$, and a  ${\IZ}_n$ subgroup that similarly has fixed points at $z_1=0$.
We get a knot provided $n$ and $m$ are relatively prime; otherwise, we get a link in $S^3$.

In any topological theory in three dimension, the partition function on $S^3$ with an $(n,m)$ torus colored by representation $R_i$ inserted can be written as follows:
\eqn\knots{
Z(S^3, K_{n,m}, R_i) = \sum_{j,k,\ell} K_{0k}{N^{k}}_{ij}{(K^{-1})^{j}}_{\ell} {S^{\ell}}_p,
}
Here, $K$ is an element of $SL(2,\IZ)$ that takes the $(0,1)$ cycle to $(n,m)$ cycle, i.e.
$$
K = \pmatrix{ a & n \cr
              b &  m \cr} \;\;  \in \;\; SL(2,{\bf Z}).
$$
with $am-nb =1$ since then the action of ${\cal O}_{R_i}^{(0,1)}$ on $|0\rangle$, and ${\cal O}^{(n,m)}_{R_i}$ on $ K|0\rangle$ agree. Any such $K$ can be written explicitly in terms of strings of $S$ and $T$ matrices
$$
S = \pmatrix{ 0 & -1 \cr
              1 &  0 \cr} , \;\; T = \pmatrix{1& 1\cr 0&1} \;\; \in \;\; SL(2,{\bf Z}).
$$
with the only ambiguity being the choice of framing $K \mapsto K T^f$ with integer $f$. For every particular $(n,m)$, eq. \knots\ is fully explicit and allows to compute straightforwardly the corresponding amplitude. For comparison with knot theory, it is most convenient to consider
the normalized amplitude, where we set the expectation value of the unknot to $1$,
$$
\frac{Z(S^3, K, \tableau{1})}{Z(S^3, \bigcirc, \tableau{1})}
$$
\centerline{\includegraphics[width=5cm]{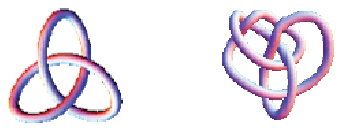}}
\noindent{\ninepoint \baselineskip=2pt {\bf Fig. 1.} {Two examples of torus knots, with winding numbers (2,3) and (4,5). }}

Introduce "knot theory" variables,
$$
{\bf t } = \sqrt{t} , \ \ \ {\bf q} = -\sqrt{q/t}, \ \ \ {\bf a}^2 = t^{N} \sqrt{t/q}.
$$
In terms of these, the $(2,3)$ torus knot (the trefoil) invariant, for example, is
$$
\frac{Z(S^3, K_{2,3}, \tableau{1})}{Z(S^3, \bigcirc, \tableau{1})} = {\bf a} {\bf q}^{-1} + {\bf a} {\bf q} {\bf t}^2 + {\bf a}^2 {\bf t}^3.
$$
For the (4,5) torus knot one gets
$$
\frac{Z(S^3, K_{4,5}, \tableau{1})}{Z(S^3, \bigcirc, \tableau{1})} =
$$
$$
={\bf a}^3 {\bf t}^4 + {\bf a}^3 {\bf q}^3 {\bf t}^6 + {\bf a}^3 {\bf q} {\bf t}^4 + {\bf a}^3 {\bf q}^{-1} {\bf t}^2 + {\bf a}^3 {\bf q}^{-3}
+ {\bf a}^4 {\bf q}^2 {\bf t}^7 +
$$
$$
+ {\bf a}^4 {\bf q} {\bf t}^7 + {\bf a}^4 {\bf t}^5 + {\bf a}^4 {\bf q}^{-1} {\bf t}^5 + {\bf a}^4 {\bf q}^{-2} {\bf t}^3 + {\bf a}^5 {\bf t}^4.
$$
Further examples are given in the Appendix.

\subsec{Relation to Knot Superpolynomials}

In the unrefined case, when $t = q$ and ${\bf t} = - 1$, the refined amplitudes reduce to the amplitudes of ordinary Chern-Simons theory. Written in terms of ${\bf q}$ and ${\bf a}$, these are known to compute the HOMFLY polynomials. Therefore for generic ${\bf t} \neq -1$, refined Chern-Simons amplitudes provide a one-parameter generalization of HOMFLY polynomials, and a natural question arises: what is the corresponding object on the mathematical side?

We conjecture that refined Chern-Simons amplitudes (in fundamental representation) compute the superpolynomial ${\cal P}(K)$ \gr\ of the knot homology theory categorifying the HOMFLY polynomial:
$$
\frac{Z(S^3, K, \tableau{1})}{Z(S^3, \bigcirc, \tableau{1})} = {\cal P}(K)({\bf a},{\bf q},{\bf t})
$$
In \AS we have shown that the conjecture holds for all $(2,2m+1)$ torus knots for any $m$, and for the $(3,3m+1), (3,3m+2)$ torus knots for $m = 1,2$, computed previously in \refs{ \gr, \rasa,\rasb}. Our more recent calculations confirm that the conjecture holds for $(3,8),(3,10),(3,11)$; $(4,5), (4,7), (4,9), (4,11)$; $(5,6), (5,8)$ and $(7,8),(7,9)$, by matching to the unpublished results of \ref\gor{Gorsky, Oblomkov Rassmussen, private communication}, made available to us by E. Gorsky.

We will give a physical explanation of these results in the next sections. We will first explain the physical origin of refined Chern-Simons theory itself, in terms of M-theory and the refined topological string.
\newsec{M-Theory Definition of Refined Chern-Simons Theory}

In this section we give a physical explanation of the above results using M-theory and the (refined) topological string. Indeed, ordinary Chern-Simons theory is known to be closely related to topological strings and also to M-theory on certain Calabi-Yau threefolds. In this section we will explain that one can turn this around and use M-theory to deduce the amplitudes of a three-dimensional topological field theory that refines Chern-Simons theory.

\subsec{Chern-Simons Theory from Topological Strings}

Let us start by summarizing the relation of ordinary Chern-Simons theory to the topological string. Topological string, or Gromov-Witten theory, on a Calabi-Yau three-fold $Y$ is given in terms of holomorphic maps $\varphi: \Sigma_g \rightarrow Y, \ {\bar \partial} \varphi = 0 $ from a Riemann surface $\Sigma_g$ into $Y$. It can be extended to open topological string, where the Riemann surface $\Sigma_g$ possibly has boundaries, whose image under the map $\varphi$ lies on a Langrangian submanifold $M \subset Y$. Since $M$ is a Calabi-Yau three-fold, $M$ is three-dimensional.

In \wst , Witten explained that, for any three-manifold $M$, the open topological string on
$$Y= T^*M,
$$
with $N$ topological D-branes on ${M}$ is the same as $SU(N)$ Chern-Simons theory on $M$. The string coupling and the level of Chern-Simons get related via
$$
g_s = {2\pi i \over k+N}.
$$
Since in this case $Y=T^*M$, there are no holomorphic curves of any kind, so only the degenerate maps $\varphi$ can contribute. These degenerate maps of the open string theory on $T^*M$ precisely reproduce the Feynman graphs of the underlying Chern-Simons theory on $M$. In particular, this implies that the Chern-Simons partition function on $M$, and the open topological string partition function on $Y$ with $N$ D-branes on $M$, $Z^{top}_{open}(T^*M)$ are the same:
$$
Z_{CS}(M) = Z^{top}_{open}(T^*M).
$$

Adding a knot $K$ to $M$, in some representation ${R}$ of the gauge group also has a topological string interpretation \ovknot, of adding D-branes wrapping a non-compact Lagrangian $L_K$. $L_K$ is a rank 2 bundle over
the knot $K$, constructed as follows.
Take a point on $K$ and the vector $V$ tangent to $K$ at that point in $M$. One obtains a two-plane in the fiber of $T^*M$ consisting of the cotangent vectors, orthogonal to $V$ in the pairing between the cotangent and tangent vectors provided by the symplectic form on $Y$ (see \refs{\ovknot,\taubes}). Such $L_K$ is topologically $\IR^2\times S^1$, and
$$L_K \cap M = K.
$$

\subsec{Chern-Simons Theory from M-Theory}

For our purposes -- to give a physical definition of the refined Chern-Simons theory -- it is important that all the above constructions are further related to M-theory with M5 branes. While in the unrefined case all sides of this relation are well known, when we consider the refined case, the M-theory will provide the sole definition of the theory.

Consider M-theory on
\eqn\background{
(Y \times TN \times S^1)_q\,,
}
where $Y$ is a Calabi-Yau manifold (for now $Y$ is arbitrary) and $TN$ is the Taub-NUT space.
The Taub-NUT space is twisted along the $S^1$, in the sense that going around the circle, the complex coordinates $z_1,z_2$ of the $TN$ rotate by
\eqn\rot{
z_1 \to q z_1,\qquad z_2 \to q^{-1} z_2,
}
so the space is not a direct product. We denoted this twist by a subscript $q$ in \background .
The M-theory partition function on this background is the same as the partition function of the closed topological string on $X$  \DVV, where one identifies $q = e^{g_s}$ with string coupling $g_s$.
To extend this to the open string \lref\cnv{
  S.~Cecotti, A.~Neitzke, C.~Vafa,
  ``R-Twisting and 4d/2d Correspondences,''
[arXiv:1006.3435 [hep-th]].
}\lref\cdv{
  M.~C.~N.~Cheng, R.~Dijkgraaf, C.~Vafa,
  ``Non-Perturbative Topological Strings And Conformal Blocks,''
[arXiv:1010.4573 [hep-th]].
} \refs{\AY, \cnv,\cdv}, we add $N$ M5 branes wrapping
$$
(M \times {\IC} \times S^1)_q
$$
where $M$ is a special Lagrangian 3-cycle in $Y.$
The branes wrap a $\;{\IC}$ subspace of $TN$ space fixed by the rotations \rot. We can take $\, {\IC}$ to correspond to the $z_1$ plane.
The partition function of the M5 branes on this background is
\eqn\mfive{
Z_{M5}(Y, M)=  {\rm Tr}\;(-1)^{F} q^{S_1-S_2}.
}
Here  $S_1$ and $S_2$ are the generators of two $U(1)_{1,2}$ rotations in \rot , and $F=2S_1$ measures the fermion number.
The M5 brane partition function \mfive\ is the same as the open topological string partition function on $Y$ with $N$ D-branes wrapping $M$,
$$
Z_{M5}(Y, M,q) = Z^{top}_{open}(Y, M,g_s),
$$
where $q=e^{g_s}$ in terms of $g_s$, the topological string coupling. In particular, in the case when $Y = T^{*} M$ with some three-manifold $M$, the partition function of M5 branes equals the $SU(N)$ Chern-Simons partition function on $M$:
$$
Z_{M5}(T^{*} M, M, q) = Z^{top}_{open}(T^{*} M, M,g_s) = Z_{CS}(M)
$$
We will now use M-theory in a slightly more general background to define what we mean by the refinement of the right hand side.

\subsec{Refined Chern-Simons Theory from M-Theory}

In certain cases, M-theory on $Y$ can be used to define a refinement of the topological string \ref\hiv{
  T.~J.~Hollowood, A.~Iqbal, C.~Vafa,
  ``Matrix models, geometric engineering and elliptic genera,''
JHEP {\bf 0803}, 069 (2008).
[hep-th/0310272].
}. Consider, as before,
M-theory on $Y\times TN \times S^1$. We fiber $TN$ over the $S^1$, so that going around the circle, the coordinates $z_1$,$z_2$ of the TN space are twisted by
\eqn\rotref{
z_1 \to q z_1,\qquad z_2 \to t^{-1} z_2.
}
We will denote the resulting space by
$$
(Y\times TN \times S^1)_{q,t}.
$$
If $t \neq q$ this alone breaks supersymmetry. However, if the Calabi-Yau $Y$ is non-compact, M-theory on $Y$ gives rise to a five dimensional gauge theory. This has an additional $U(1)_R \subset SU(2)_R$ symmetry, and supersymmetry of the theory can be preserved provided, as one goes around the $S^1,$ one performs an additional $R$-symmetry twist.

Now consider adding $N$ M5 branes on
$$(M\times \,{\IC}\times S^1)_{q,t},$$
where $\,{\IC}$ corresponds to the $z_1$ plane.  The M5 brane configuration automatically preserves a $U(1)_1 \times U(1)_2$ symmetry rotating the $z_1$ and $z_2$ planes. For it to have the $U(1)_R$ symmetry as well, an additional requirement is needed. Let us focus for now on the main case of interest for us, when
$$
Y = T^*M.
$$
The theory has the $U(1)_R$ symmetry provided $M$ admits a free $U(1)$ action. This $U(1)$ symmetry of $M$ is itself not an $R$-symmetry, however, it can be used to construct one \AS. More precisely, the action only needs to be semi-free: this corresponds to allowing a discrete subgroup of $U(1)$ to act with fixed points. This implies that $M$ is a Seifert three-manifold -- an $S^1$ fibration over a genus $g$ Riemann surface $\Sigma_g$,
$$
S^1 \rightarrow M \rightarrow \Sigma_g
$$
where the $U(1)$ action comes from the rotation of the fiber. The ordinary Chern-Simons theory on Seifert spaces was studied recently in \ref\bw{
  C.~Beasley, E.~Witten,
  ``Non-Abelian localization for Chern-Simons theory,''
J.\ Diff.\ Geom.\  {\bf 70}, 183-323 (2005).
[hep-th/0503126].
}\ref\BeasleyMB{
  C.~Beasley,
  ``Localization for Wilson Loops in Chern-Simons Theory,''
[arXiv:0911.2687 [hep-th]].
}\ref\BeasleyHM{
  C.~Beasley,
  ``Remarks on Wilson Loops and Seifert Loops in Chern-Simons Theory,''
[arXiv:1012.5064 [hep-th]].
}.

With the additional $U(1)_R$ symmetry preserved, the M5 brane partition function on $M \times {\IC} \times S^1$ defines an index,
\eqn\rindex{Z_M(T^*M, q,t)=   {\rm Tr}\, (-1)^{F}\, q^{S_1-S_r}\, t^{S_r-{S}_2}.
}
Note that both $S_2$ and $S_r$ are R-symmetries, as they correspond to rotations in the normal bundle to the M5 branes. Their difference $S_2-S_r$ acts as a global symmetry, as the three dimensional ${\cal N}=2$ theory at low energies has a unique $U(1)_R$ symmetry. A pair of the supercharges $Q, {\bar Q}$ preserved by the brane, whose
$(S_1,S_2,S_r)$ charges are $\pm(1/2, 1/2, 1/2)$ can be used to define the above index, see for example \ref\gaiotto{
  L.~F.~Alday, D.~Gaiotto, S.~Gukov, Y.~Tachikawa, H.~Verlinde,
  ``Loop and surface operators in N=2 gauge theory and Liouville modular geometry,''
JHEP {\bf 1001}, 113 (2010).
[arXiv:0909.0945 [hep-th]].
}. The index localizes on configurations that are annihilated by $Q, {\bar Q}$. Moreover, for $q=t$ it reduces to the unrefined index \mfive.
We will take this index \rindex\ as the definition of the refined $SU(N)$ Chern-Simons partition function on $M$:
\eqn\mcs{
Z_{M}(T^*M,q,t) \equiv Z_{CS}(M, q,t).
}
This is analogous to the unrefined case, where the $N$ M5 brane partition function on $M$ in $T^*M$ equals the ordinary $SU(N)$ Chern-Simons partition function on $M$. Unlike in the unrefined case, we now do not have an alternative definition of the theory.

We can also include knot observables in the refined Chern-Simons theory. As we explained above, in the ordinary topological string, including a Wilson loop $K$ on a three manifold $M$ in Chern-Simons theory corresponds to adding D-branes on a special Lagrangian $L_K$ in $T^*M$. To extend this to the refined case, both the knot $K$ and the three-manifold $M$ have to respect the $U(1)_R$ symmetry that is needed to define the theory. As explained in \AS, this implies that the allowed knots and links in $M$ are the Seifert knots \BeasleyHM, wrapping the $S^1$ fibers over $\Sigma_g$ in $M$. In particular, in the simplest case $M = S^3$ where most of the actual calculations have been performed so far, the allowed knots and links are presicely the torus knots and links.

\subsec{The extra $U(1)$ Symmetry}

Given the importance of the existence of the extra $U(1)$ symmetry for the definition of the index, let us elaborate this point. The $U(1)$ symmetry we need is constructed as follows. $M$ being a Seifert three manifold has a nowhere vanishing vector field $V$. Thinking of $M$ as an $S^1$ fibration over a Riemann surface, $V$ acts by rotating the $S^1$ fiber.  Using $V$, we can define at a rank-two sub bundle of the cotangent bundle $T^*M$ to $M$. At each point in $M$ the fiber of this bundle consists of those cotangent vectors that are orthogonal to $V$. The $U(1)$ action we want to use, whose generator we called $S_R$ above, corresponds to rotation of the fibers of this bundle.

To be sure we can use this $U(1)$ action to define the refined index, we need show that $T^*M$ admits a metric that has this $U(1)$ action as an isometry, and moreover, that this metric preserves supersymmetry. We will give now argue that such a metric exists\foot{We thank E. Witten for discussions of this point.}. Consider, to begin with, $N$ M5 branes on a three-manifold of the special form, $M = S^1 \times \Sigma $, times the flat ${\IR}^{2,1}$, inside $T^*M \times {\IR}^{4,1} = T^*\Sigma \times T^*S^1 \times {\IR}^{4,1}.$ The theory has
$SO(2)_1\times SO(2)_2\times SO(2)_R$ as a subgroup of the symmetry group. Here, $SO(2)_R$ rotates the fibers of $T^*\Sigma,$ $SO(2)_1$, is a part of the Lorentz group of the brane, and $SO(2)_2$ rotates the normal directions to the brane in ${\IR}^{4,1}$. $SO(2)_2$ and $SO(2)_R$ are both $R$-symmetries of the ${\cal N}=4$ supersymmetric theory in three dimensions on the brane. It is helpful to reduce the $(2,0)$ theory along the $S^1$ (viewing this as the M-theory circle). Then, the theory we are discussing is obtained by compactifying five dimensional Yang-Mills theory on $\Sigma$, to obtain the three dimensional Yang-Mills theory on  ${\IR}^{2,1}$.

Now consider M5 branes on a Seifert three manifold obtained by fibering the $S^1$ non-trivially over $\Sigma$. In this case,
half of the supersymmetry of the theory is broken: the theory on ${\IR}^{2,1}$ has ${\cal N}=2$ supersymmetry in three dimensions.
This also corresponds to the fact that $T^*M$ is an honest Calabi-Yau three-fold. Supersymmetries that get preserved by the background have their $SO(2)_{2}\times SO(2)_R$ charges correlated. We can take the supercharges that survive to have $S_2=S_R$. Thus, one combination of $S_2$ and $S_R$ becomes an $R$ symmetry of the theory. In addition, we potentially get a global $U(1)$ symmetry, corresponding to $S_2-S_R$, if both $S_2$ and $S_R$ survive as the symmetries. If this is the case,
we can define the refined index \rindex\ of the M5 brane theory on $M\times {\IR}^2 \times S^1.$

What is the effect of fibering the $S^1$ over $\Sigma$ on the M5 brane theory? We can answer this question from the perspective of the $(2,0)$ theory or better yet in its dimensionally reduced version -- where we reduce on the $S^1$ fiber of the Seifert three manifold, as we did above. The answer is that giving the $S^1$ bundle over $\Sigma$ a degree $p$ corresponds to turning on an ${\cal N}=2$ Chern-Simons coupling on ${\IR}^{2,1}$,
\eqn\cherns{ p \int_{\IR^{2,1}}d^4\theta \;  {\rm Tr} {\cal V} \Sigma({\cal V}) .}
 where ${\cal  V}$ is the ${\cal N}=2$ vector multiplet, and $ \Sigma({\cal V}) = \epsilon^{\alpha \beta} {\bar D}_{\alpha} D_{\beta}{\cal V}$ is the linear superfield \ref\Seiberg{
  O.~Aharony, A.~Hanany, K.~A.~Intriligator, N.~Seiberg and M.~J.~Strassler,
  ``Aspects of N=2 supersymmetric gauge theories in three-dimensions,''
Nucl.\ Phys.\ B {\bf 499}, 67 (1997).
[hep-th/9703110].
}. The crucial point is that this coupling is neutral under both $SO(2)_2$ and $SO(2)_R$. Thus, turning it on breaks neither symmetry, and both $S_2$ and $S_R$ survive as generators of R-symmetries in the theory on ${\IR}^{2,1}$. To see that the ${\cal N}=2$ Chern-Simons coupling preserves both symmetries, it suffices to note that one of its terms is the bosonic Chern-Simons coupling $p \int_{{\IR}^{2,1}}\omega_{CS}(A)$, where $A$ is the gauge field on ${\IR}^{2,1}$,  clearly neutral under both of the symmetries, and $\omega_{CS}(A) = {\rm Tr} AdA - \frac{2}{3}A^3$ is the Chern-Simons three form.

The origin of the Chern-Simons coupling is purely topological. The fastest way to see that is to recall that, viewing the circle fiber to $M$ as the M-theory circle, the five dimensional YM theory is a theory on $N$ D4 branes, and that has a term
\eqn\wz{\int_{\Sigma\times {\bf R}^{2,1}} F_{RR} \wedge\omega_{CS}(A)}
in its action. The 11-dimensional interpretation of $F_{RR}$ is the curvature of the circle bundle corresponding the M-theory circle.
In our present case, we would have
$$
\int_{\Sigma} F_{RR} = p
$$
resulting in the above coupling. The rest of \cherns\ is fixed by ${\cal N}=2$ supersymmetry. It may seem that we are using some very particular facts about the couplings on the D4 branes to argue this. This is not the case. The origin of the term in \wz\ is a the fact that the $(2,0)$ theory has a self-dual two-form tensor on it. In any attempt to write down the action for the corresponding theory there is a peculiar Wess-Zumino type term that arizes -- albeit involving the metric on the six manifold (see for example
\ref\AganagicZK{
  M.~Aganagic, J.~Park, C.~Popescu and J.~H.~Schwarz,
  ``Dual D-brane actions,''
Nucl.\ Phys.\ B {\bf 496}, 215 (1997).
[hep-th/9702133].
}\ref\AganagicZQ{
  M.~Aganagic, J.~Park, C.~Popescu and J.~H.~Schwarz,
  ``World volume action of the M theory five-brane,''
Nucl.\ Phys.\ B {\bf 496}, 191 (1997).
[hep-th/9701166].
}\ref\SchwarzMC{
  J.~H.~Schwarz,
  ``Coupling a selfdual tensor to gravity in six-dimensions,''
Phys.\ Lett.\ B {\bf 395}, 191 (1997).
[hep-th/9701008].
}). Using dimensional reduction to get to a five dimensional Yang-Mills the term \wz\ arizes, with $F_{RR}$ as the curvature of the $S^1$ bundle.

Thus, in the specific case when $M$ is a Seifert three manifold, the theory has, in addition to a $U(1)_R$ symmetry $S_2$, an additional $U(1)$ symmetry generated by $S_R$. Correspondingly, when $M$ is a Seifert three manifold, we can define the refined index \rindex .

\subsec{Related Work}

In \wr\  a physical approach to knot homology was proposed, based on studying gauge theory on D4-branes wrapping a four-manifolds with a boundary on the three manifold $M$, where the knots live, times a thermal $S^1$ (there were other duality frames studied in \wr\ as well, but we will focus on this one, as it is closest to us).  The advantage of the approach initialized in \wr\ and developed further in \wrg\wrw\ is that it provides one a way to get at knot homology groups themselves, not relying on indices that exist when $M$ is special. But, nevertheless it is important to note that the physical setting of \wr\ and the one we use here are related by a simple duality.

To define the refined Chern-Simons theory on a three manifold $M$, we needed to study M-theory on $Y\times TN \times S^1$, where $Y = T^*M$ with $N$ M5 branes on $M\times {\IC}\times S^1$. Consider a dual description of this, by dimensionally reducing on the $S^1$ of the Taub-Nut space. Without M5 branes, we would obtain IIA string theory on the geometry,
$$
Y\times {\IR}^{3} \times S^{1}
$$
with a D6 brane wrapping  $Y \times S^{1}$ and sitting at the origin of ${\IR}^{3}.$   Adding the $N$ M5 branes on $M\times {\,\IC}\times S^1$, we get IIA string theory with the addition of $N$ D4 branes, wrapping $M\times {S^1}$ times a half-line ${\IR}_+$ in ${\IR}^3$, ending on the D6 brane. This is a D4 brane on a four manifold ${\IR}_+\times M\times S^1$, with the specific boundary condition imposed by the D6 brane. This setup is the same as that in \wr\ (see the discussion on the bottom of p. 13 of \wr\ and else where in the paper). Now consider how the symmetry generators map between the two pictures. From this we will deduce the index in IIA corresponding to the refined Chern-Simons partition function computed in M-theory, and recover in the unrefined limit, the index computation in \wr\wrg\wrw\ that gave rise to the Jones polynomial.

Before we add branes, the Taub-NUT geometry has an $SU(2)_{\ell} \times SU(2)_{r}$ isometry. We used the $U(1)_{\ell}\times U(1)_r$ subgroup of it in the definition of the index. The $U(1)_{\ell} \times U(1)_{r}$ act one the complex coordinates $(z_1, z_2)$ of the TN space by $(e^{i (\theta_{\ell} + \theta_r)/2}z_1, e^{i (-\theta_{\ell} + \theta_r)/2}z_2).$  Asymptotically, Taub-NUT looks like $S^{1} \times {\IR}^{3}$ and the $U(1)_{\ell}$ isometry rotates the $S^{1}$, while the $SU(2)_{r}$ rotates the base geometry.  So upon dimensional reduction, the charge under $U(1)_{{\ell}}$ becomes the D0-brane charge, while the charge under $SU(2)_{r}$ becomes the spin in the base ${\IR}^{3}$.
In addition to the this, IIA and M-theory have a common $SU(2)_R$ R-symmetry of a five-dimensional gauge theory.

The we add branes preserve the $U(1)_r$ subgroup of the $SU(2)_r$ rotation group, for any $M$. For any $M$, setting $q=t=q_0$, the partition function of the M5 brane theory \mfive\ equals the partition function of the D4 brane theory in this background
$$
Z_{D4}(T^*M, q_0)=   {\rm Tr}\, (-1)^{F}\, q_0^{Q_0}
$$
and both equal to the partition function of the ordinary $SU(N)$ Chern-Simons theory on $M$. In the D4 brane context, this was shown in \wr\ and studied further in \wrg  . In \wr\wrg\ the Chern-Simons level arizes due to non-zero value of the Wilson line of the RR 1-form potential $C$ in IIA string theory, $\int_{S^1} C$. This couples to D0 brane charge. It is the same as $\log q_0$, the chemical potential for the D0 branes turned on in our setting.

When $M$ is a Seifert three manifold both the M5 brane, and the D4 brane theories should also preserve a $U(1)$ subgroup of the $SU(2)_R$-symmetry group of the five dimensional background, by the duality. Then, we can define the refined index \rindex , giving rise to the refined Chern-Simons theory, and depending on one more parameter.
The refined partition function \rindex\ becomes the partition function of the theory on $N$ D4 branes in this background
$$
Z_{D4}(T^*M, q_0,y)=   {\rm Tr}\, (-1)^{F}\, q_0^{Q_0}\, y^{2 J_3 -2{S}_R}.
$$
Here $q_{0}=\sqrt{qt}$, $y=\sqrt{q/t}$, $Q_{0}$ is the D0 brane charge, and $J_{3}$ is the generator of the rotation group in
${\IR}^{3}$, and $S_R$ is the generator of the $U(1)$.

We could have also studied a different circle reduction, where we reduce  to IIA on the thermal $S^1$ instead. In this case, we get IIA string theory on $Y\times TN$, with the four dimensional omega background. Adding $N$ M5 branes, maps to adding $N$ Dbranes wrapping $M\times\; \IC$. The D4 branes are nothing but surface operators, in the four ${\cal N}=2$  gauge theory in omega background (more precisely five dimensional gauge theory on $TN \times S^1$  where the data of the gauge theory is determined by the choice of $Y$ in the usual way. The surface operators in 4d gauge theories were studied extensively in
\ref\gwsurf{S.Gukov and E.Witten, "Rigid Surface Operators",arXiv:0804.1561}\GGV\ref\GaiottoFS{
  D.~Gaiotto,
[arXiv:0911.1316 [hep-th]].
}\ref\KozcazAF{
  C.~Kozcaz, S.~Pasquetti and N.~Wyllard,
JHEP {\bf 1008}, 042 (2010).
[arXiv:1004.2025 [hep-th]].
}\ref\AldayVG{
  L.~F.~Alday and Y.~Tachikawa,
Lett.\ Math.\ Phys.\  {\bf 94}, 87 (2010).
[arXiv:1005.4469 [hep-th]].
}\ref\dgl{
  T.~Dimofte, S.~Gukov and L.~Hollands,
  ``Vortex Counting and Lagrangian 3-manifolds,''
Lett.\ Math.\ Phys.\  {\bf 98}, 225 (2011).
[arXiv:1006.0977 [hep-th]].
}\ref\ltsurf{
M.Taki,
"Surface Operator, Bubbling Calabi-Yau and AGT Relation",
JHEP 1107 (2011) 047,
arXiv:1007.2524
}\ref\ksurf{H. Awata, H. Fuji, H. Kanno, M. Manabe, Y. Yamada, "Localization with a Surface Operator, Irregular Conformal Blocks and Open Topological String",arXiv:1008.0574}.

\newsec{Solving the Refined Chern-Simons Theory via M-Theory}

Having used M-theory to define the refined Chern-Simons theory on Seifert three-manifolds $M$, we will explain how to solve it, and obtain the results quoted in section 2. The basic idea is to use topological invariance of the theory to solve it on simple pieces first, and recover the rest by gluing. The key amplitude we will obtain in this way corresponds to taking
$$
M_L = S^1 \times {\IR}^2
$$
inside
$$
Y_L=T^*M_L = T^*(S^1 \times{\IR}^2) = {\IC}^* \times \,{\IC}^2.
$$
The free $U(1)$ action on $M_L$ corresponds to the rotation of the $S^1$, and the $U(1)_R$ symmetry one gets from it acts by rotating the fiber in $T^* {\IR}^2$. The Calabi-Yau $Y_L={\IC}^* \times \,{\IC}^2$ is essentially a flat space, hence, non-compact; because of this, the $U(N)$ symmetry of the theory on $M_L$ is a global symmetry. The partition function depends on the values of scalars $x_I$, $I=1, \ldots N$ of the background linear multiplets. These have interpretation as the positions of M5 branes along the fiber of $T^* S^1$ (this is a cylinder, and $x_I$ parameterize the linear direction). We can, alternatively, view $M_L$ as a solid torus $S^1 \times D$ where the path integral computes a wave function depending on the boundary conditions we impose -- by topological invariance the two viewpoints are equivalent.
With the branes widely separated, computing the partition function of the M5 brane theory on $(M_L \times {\IC} \times S^1)_{q,t}$ inside $(Y_L \times TN \times S^1)_{q,t}$ amounts to counting BPS states of M2 branes stretching between them -- i.e., cohomologies of the moduli of holomorphic curves embedded in Y and with boundaries on the M5 branes. As shown in detail in \AS, this spectrum is very simple and consists precisely of two cohomology classes. (From the 3D perspective, the contributions to the index come from W-bosons and the adjoint scalar, parameterizing the position of the M5 branes along the $\;{\IC}^2$ directions transverse to its world volume).

\centerline{\includegraphics[width=5cm]{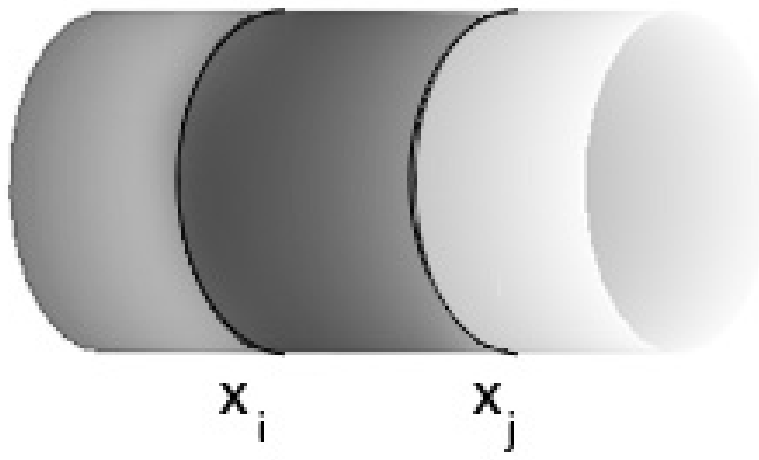}}
\noindent{\ninepoint \baselineskip=2pt {\bf Fig. 2.} {The D-branes wrap two $S^1$'s on the cylinder ${\IC}^*$ in $Y_L$. The  open topological string is counting the maps to the annulus between them.  }}

Because of this, one can compute the partition function explicitly as a function of the positions of the M5 branes. One finds
\eqn\partfunc{
Z(S^1 \times {\IR}^2)(q,t) =\prod_{1\leq I<J \leq N}\prod_{n=1}^{\infty}\frac{ (q^{-n/2} e^{(x_J - x_I)/2} - q^{n/2} e^{(x_I-x_J)/2})}
{ (t^{-1/2} q^{-n/2} e^{(x_J - x_I)/2} - t^{1/2} q^{n/2} e^{(x_I-x_J)/2})}
}
Note that this is consistent with the known result for the unrefined case: considering $N$ branes and symmetrizing, for $q=t$ we recover
\eqn\wavefunc{
Z(S^1 \times {\IR}^2)(q,q) = \prod_{1\leq I<J \leq N} (e^{(x_J - x_I)/2} - e^{(x_I-x_J)/2}),
}
which is the Chern-Simons partition function on a solid torus \ems\  $S^1 \times {\IR}^2$. To avoid dealing with infinite products, we can put
\eqn\betin{
 t = q^{\beta}, \qquad \beta \in \IN,
}
where $\beta$ is a positive integer. This specialization \betin\ is inessential -- while it will simplify the intermediate formulas, it is easy to reinstate arbitrary $\beta$ by inspection. At integer $\beta$, the partition function \partfunc\ can be written as a finite product:
\eqn\wavefuncref{
Z(S^1 \times {\IR}^2)(q,t) = \prod_{m=0}^{\beta -1} \prod_{1\leq I<J \leq N}\; (q^{-m/2} e^{(x_J-x_I)/2}- q^{m/2} e^{(x_I-x_J)/2}) \equiv \Delta_{q,t}(x).
}
This is a partition function of M-theory on $T^{*}(S^1 \times {\IR}^2)$. At the same time, by \mcs , this should be regarded as the partition function of refined Chern-Simons theory on the solid torus. This simple result allows to compute the partition functions for more complicated 3-manifolds $M$ by cutting and gluing.

To consider gluing, it is useful to think of this in terms of the quantization of Chern-Simons theory. In this picture, as explained in Sec. 2, Chern-Simons theory on a manifold with $T^2$ boundary has a Hilbert space of states labeled by representations $R$ of $SU(N),$ satisfying a certain condition. Then, the partition function eq. \wavefunc\ is interpreted as the wavefunction of the vacuum state. Our results on refined Chern-Simons theory seem to suggest that at $q \neq t$, the theory has the same Hilbert space of states, but the wavefunction of the vacuum becomes \wavefuncref\ :
$$
\big< 0 \big| x \big> = \Delta_{q,t}(x)
$$
Given the form of this vacuum, the most natural choice of  basis in the Hilbert space is the one that has wavefunctions given by Macdonald polynomials:
$$
\big< R_i \big| x \big> = M_{R_i}(e^x) \Delta_{q,t}(x)
$$
This is because Macdonald polynomials are the orthogonal polynomials associated with the $\Delta_{q,t}(x)$ measure, i.e. in Macdonald basis the identity operator has a diagonal matrix:
$$
\big< R_i \big| R_j \big> =  \int dx_1 \ldots dx_N \Big| \Delta_{q,t}(x)\Big|^2 M_{R_i}(e^x) M_{R_j}(e^{-x}) = g_i 1_{ij}
$$
where $g_i = ||M_{R_i}||^2$ is the Macdonald integral norm.

Now we can take two copies of $T^2$ with opposite orientation and glue them with trivial identification of the boundary, what gives the partition function of refined Chern-Simons theory on $S^1 \times S^2$ and $M$-theory on $T^{*}(S^1 \times S^2)$. This corresponds to taking the overlap of the two solid toruus wave functions, by integrating over $x_1, \ldots, x_N$:

$$
Z(S^1 \times S^2) = \big< 0 \big| 0 \big> = \int dx_1 \ldots dx_N \Big| \Delta_{q,t}(x)\Big|.^2
$$
Instead of considering $S^1 \times S^2$, one can consider more general fibrations with the first Chern class $p$, by multiplying one of two wave functions by

$$
e^{- p \sum\limits_{i} \frac{x_i^2}{2} }
$$
before gluing. For example, $S^3$ is a Hopf fibration, with $p = 1$. Accordingly, the partition function of refined Chern-Simons theory on $S^3$ is given by the integral

\eqn\zsthree{
Z(S^3) = \int dx_1 \ldots dx_N \Big| \Delta_{q,t}(x)\Big|^2 e^{- \sum\limits_{i} \frac{x_i^2}{2} } = \big< 0 \big| TST  \big| 0 \big>
}
i.e., the vacuum matrix element of the operator TST. Similarly, one deduces that other matrix elements of this operator are given by integrals

$$
\big< R_i \big| T S T \big| R_j \big> = \int dx_1 \ldots dx_N \Big| \Delta_{q,t}(x)\Big|^2 e^{- \sum\limits_{i} \frac{x_i^2}{2} } M_{R_i}(e^x) M_{R_j}(e^{-x})
$$
The fact that this takes the explicit form that we gave before \nsm\ , \ntm\ was proven in the 90's in the context of proving Macdonald Conjectures.

Other examples of gluing include the Verlinde numbers that compute the amplitude on $S^1 \times S^2$ with three Wilson lines wrapping the $S^1$:
$$
Z(S^1 \times S^2| R_i, R_j, R_k) = \int dx_1 \ldots dx_N \Big| \Delta_{q,t}(x)\Big|^2 M_{R_i}(e^x) M_{R_j}(e^{x}) M_{R_k}(e^x) = N_{ijk}
$$
The fact that they satisfy the Verlinde formula \vftwo\ then follows from three-dimensional topological invariance -- see the Appendix to \AS \ for derivation.

\newsec{Large-N Dual of Refined Chern-Simons Theory}

Ordinary Chern-Simons theory gets related, by string dualities, to other problems in mathematics and physics. Particularly important is that $SU(N)$ Chern-Simons theory has a large $N$ dual description in terms of the refined topological string on a different manifold.
Among other applications, this large $N$ duality was used in \refs{\ovknot,\gsv} to physically explain the integrality of coefficients of knot invariants.
In this section we will first review this duality in the ordinary unrefined case, and then explain how it extends to the refined theory.

\subsec{Large $N$ Duality in Ordinary Chern-Simons Theory}

Recall, from section 2, that $SU(N)$ Chern-Simons theory on $S^3$ is the same as the open topological string on
$$Y=T^*S^3,$$
with $N$ D-branes on the $S^3$.
Gopakumar and Vafa showed this has a large $N$ dual, the ordinary topological string theory on
$$
{X} = {\cal O}(-1) \oplus {\cal O}(-1) \rightarrow {\IP}^1.
$$
The duality is a large $N$ duality in the sense of 't Hooft \ref\thooft{G. t'Hooft, "Planar Diagram Theory for Strong Interactions", Nucl. Phys. 72 (1974) 461.}. The topological string coupling $g_s$ is the same on both sides -- it is related to the level $k$ in Chern-Simons theory by
$$g_s = {2\pi i \over k+N}.$$
The later is the effective coupling constant of Chern-Simons theory, due to the famous shift of
$k\rightarrow k+N$, generated by quantum effects. The rank $N$ of the gauge theory is related to the area of the ${\IP^1}$ in $X$ by
$\lambda = e^{-Area({\IP^1})}$ by
\eqn\llambda{\lambda = q^N,
}
 where $q = e^{g_s}.$
This is a large $N$ duality, since when the geometry of $X$ is classical, $\lambda$ is continuous, and this is only true in the limit of large $N$. The duality in this case also has a beautiful geometric interpretation: it is a geometric transition that shrinks the $S^3$ and  grows the ${\IP^1}$ at the apex of the conifold, thereby taking $Y$ to $X$ \gvI. The duality has been checked, at the level of partition functions,  to all orders in the $1/N$ expansion \gvI .
\lref\AV{
  M.~Aganagic and C.~Vafa,
  ``Mirror symmetry, D-branes and counting holomorphic discs,''
[hep-th/0012041].
}

As we explained in section $2$, adding knots on the $S^3$ corresponds to introducing non-compact branes on a Lagrangian $L_K$ in $Y$, intersecting the $S^3$ along a knot $K$,
$$
K = L_K \cap S^3.
$$
The geometric transition affects the interior of $X$ and $Y$, but not their asymptotics, which are the same.
A non-compact Lagrangian $L_K$ on $Y$, goes through the transition to give a Lagrangian $L_K$ on $X$.
The correspondence of Chern-Simons theory with knots and the open topological string on $X$ has been proven for the case of the unknot and the hopf link, colored by arbitrary representations \refs{\ovknot, \AV, \akv, \amv}. This was also recently extended to toric knots in
\ref\BriniWI{
  A.~Brini, B.~Eynard and M.~Marino,
  ``Torus knots and mirror symmetry,''
[arXiv:1105.2012 [hep-th]].
}.

\subsec{Large $N$ Duality in Refined Chern-Simons Theory}

The large $N$ duality relating open topological string on $Y=T^*S^3$, i.e, Chern-Simons theory on the $S^3$ to topological string on $X= {\cal O}(-1) \oplus {\cal O}(-1) \rightarrow {\IP}^1$ was expected to extend to the refined topological string \refs{\civ,\gsv, \toda}, where we view refinement as a deformation\foot{We do not expect the large $N$ duality to extend to the full suprestring theory as in the unrefined case. Instead, we expect it to be the case for the topological, i.e. BPS sector of the theory, captured by the index.}.
With the refined Chern-Simons theory in hand, we can offer compelling evidence that the refined theory indeed inherits the large $N$ duality. The simplest test of the duality is to compare partition functions (without knot observables) on two sides of the duality.

On $X$, we are computing the partition function of the refined closed topological string, which is well known. Let $\lambda$ parameterize the size of the ${\IP}^1$ in $X$ as  $\lambda =e^{-Area({\IP}^1)}$, were by the area we really mean the mass of a BPS state wrapping the $\IP^1$. The partition function of the closed topological string in this background is  \civ\
\eqn\refc{
Z^{top}(X, \lambda, q,t) = \exp\Bigl(-\sum_{n=0}^{\infty} {\lambda^n \over n (q^{n/2} - q^{-n/2})(t^{n/2} - t^{-n/2})}\Bigr),
}
up to simple factors that correspond to classical intersection numbers on the Calabi-Yau, and are ambiguous. Here $q$ and $t$ are the parameters defining the omega-background,  $q=e^{\epsilon_1}$, $t = e^{-\epsilon_2}$. On $Y$, we are computing the partition function of $N$ M5 branes, that is given by the Macdonald-type integral in \zsthree ,
\eqn\lncs{
Z(S^3, N,q,t) = \int dx_1 \ldots dx_N \big| \Delta_{q,t}(x)\big|^2 e^{- \sum_{i} \frac{x_i^2}{2 g_s} }
}
 The value of the integral is known explicitly for arbitrary finite $N$:
\eqn\lncsa{
Z(S^3, N,q,t)  = \prod_{m=0}^{\beta -1} \prod_{i=1}^{N-1}(1-t^{N-i } q^{m})^{i},
}
where $q = e^{g_s}$ and $t = q^{\beta}$. Taking the large $N$ limit of this exact finite-$N$ expression, we find (see \AS \ for a derivation) that \lncs\ and \refc\ agree,
$$
Z(S^3,N,q,t) = Z^{top}(X, \lambda,q,t),
$$
provided we identify
\eqn\map{\lambda =  t^{N} \sqrt{t/q}.
}
This generalizes the relation
between the $\lambda$ and $N$ in the unrefined case.
The fact that $N$ $\ep_1$-branes on the $S^3$ back-react on the geometry in such a way to open up a ${\IP}^1$ of size $N\ep_2$, is as expected on general grounds, as discussed recently in detail in \acdkv . By $\ep_1$ branes we mean the branes wrapping the $z_1$ plane rotated by the parameter $q=e^{\ep_1}$, and by $\ep_2$ branes, the branes wrapping the $z_2$ plane, rotated by $t = e^{\ep_2}$. The factors $\sqrt{t/q}$, which vanish in the unrefined theory at $q=t$, correspond to quantum shifts of the moduli by $(\ep_2-\ep_1)/2,$ and are typical \refs{\toda,\acdkv}.

\subsec{Large $N$ Duality and Integrality of Knot Invariants }

Two parallel approaches were developed to integrality of knot invariants. The one coming from mathematics, due originally to Khovanov, introduced the idea of knot homology, revealing the fact that Chern-Simons knot invariants are actually Euler characteristics of certain complexes of vector spaces \kh. On the physical side, nearly simultaneously with Khovanov's work, an explanation for this phenomenon was put forward in \ovknot .

The physical explanation from \ovknot\ was based on the two duality relations, the large $N$ duality relating $SU(N)$ Chern-Simons theory on the $S^3$ to the topological string on $X = {\cal O}(-1)\oplus {\cal O}(-1)\rightarrow \IP^1$, and the duality of the topological string on $X$ with M-theory on $(X \times TN\times S^1)_q$. In \ovknot\ the authors showed that, taken together, the two dualities imply that computing HOMFLY invariant of a knot $K$ in $S^3$ is related to the index \mfive\ of M5 branes wrapping $(L_K\times \IC\times S^1)_q$,
\eqn\largenknots{
Z_{CS}(K, S^3, N, q) = Z_M(L_K, X, \lambda, q).
}
The right hand side is the partition function of M5 branes wrapping the Lagrangian $L_K$ in $X$, associated to the knot $K$. This partition function is computed by the index\foot{For simplicity, we focus on knots colored by fundamental representation, so as to suppress an additional chemical potential one would have needed otherwise, to keep track of representations.}
\eqn\parti{
Z_M(L_K,X, \lambda, q) =   {\rm Tr}_{{\cal H}_{BPS}} (-1)^{F} q^{S_1-S_2} \, \lambda^Q ,\,
}
counting the BPS states of M2 branes ending on the M5 branes on $L_K$. The grade $\lambda$ of HOMFLY gets interpreted as the Kahler parameter $\lambda$, of the large $N$ dual \llambda.
%
%
%
%
The fact that the knot invariants get related to a problem of counting BPS particles explains their integrality.

A natural further conjecture, made in \gsv , is that the spaces of BPS states of M2 branes ending on $L_K$ in $X$, where one keeps track of the spins $S_1$ and $S_2$ separately, are precisely the homologies of the knot $K$.
Later, in \gr\ motivated by this conjecture, a candidate for the right knot homology theory to match the physics was put forward. There, the authors proposed existence of a triply graded knot homology theory categorifying the HOMFLY polynomial.  The Poincare polynomial of this theory has been named the superpolynomial in \gr .
In terms of M-theory, the \gsv\ conjecture implies that the superpolynomial of \gr\ is computed by counting the dimensions of spaces of BPS states on $X$, keeping track of their $(S_1,S_2)$ spins:
%
by refining the counting in \parti\ to
\eqn\supp{
 {\rm Tr}_{{\cal H}_{BPS}}  (-1)^{F} q^{S_1} t^{-S_2} \, \lambda^Q
 }
Note that, since $F$, the fermion number also equals $2S_1$, the $(-1)^F$ can be absorbed into $q^{S_1}$, to make all the coefficients in \supp\ positive. Is important to emphasize that, while one can define \supp\ for any knot $K$, this is not the partition function of the M5 branes on $L_K$, in any way of defining it. The partition function would need to extend over the whole Hilbert space ${\cal H}$, a subspace of which is the space of BPS states, ${\cal H}_{BPS}$. This would receive contributions from non-BPS states as well, since it is not an index. Equivalently, there should be no path integral way of computing it \supp , analogously to what Chern-Simons theory does for the Jones polynomial\foot{We are grateful to Edward Witten for discussions and explanations tied to this point.}.

If however, the theory has an additional $U(1)_R$ symmetry, as when $K$ is a torus knot, than one can define a refined index:
\eqn\trui{
Z_{M}(L_K, X, q,t) =   {\rm Tr}_{{\cal H}_{BPS}} (-1)^{F} q^{S_1-S_R} t^{S_R-S_2}\, \lambda^Q ,\,
}
computed by the partition function of M5 branes on $(L_K \times \,{\IC}\times S^1)_{q,t} \ \subset \
(X\times TN\times S^1)_{q,t}$,
with the $U(1)_R$ symmetry twist as one goes around the $S^1$, to preserve supersymmetry.
The index \trui\ is counting BPS states of M2 branes on $X$, and should, equal the refined Chern-Simons knot invariant
if the large $N$ duality holds,
$$
Z_{CS}(K, S^3, N, q, t) = Z_{M}(L_K, X, \lambda, q,t).
$$

The results we presented in section two, provide support for the conjecture of \gsv\ and large $N$ duality. Recall, we showed that, starting with the refined Chern-Simons invariant of a torus knot $K$, and setting
$$
\lambda = - t^N \sqrt{t/q},
$$
we obtained the corresponding superpolynomial of $K$, at least in a large number of cases. The change of variables is exactly what corresponds to Kahler modulus of $X$, by large $N$ duality. Moreover, the refined index on $X$ and the Poincare superpolynomial agree, assuming $S_R$ in \trui\ vanishes (or is otherwise correlated with $Q$) for all BPS states of M2 branes ending on $L_K$ in $X$, where $K$ is a toric knot.
In this particular case, the Poincare polynomial and the index agree, as we found in section 2.5.

Moreover, the results of \gr\ provide a mathematical formulation of the index refined Chern-Simons theory is computing at finite $N$.
In the categorification of \gr\ an important role is played by the so called $d_N$ differential.
The $d_N$ operator is crucial in relating the HOMFLY knot homology of \gr\ and the better established $SL_N$ knot homology of Khovanov and Rozansky \kro\krt, that categorifies $SU(N)$ Chern-Simons knot invariants: as explained in \gr\ the two are related in a non-trivial way at low enough $N$. At low enough $N$ the HOMFLY homology is "too big": the $SL_N$ homology emerges from the HOMFLY knot homology only upon taking the cohomology with respect to $d_N$.

On the other hand, to relate the superpolynomial ${\cal P},$ i.e. the Poincare polynomial of the HOMFLY knot homology to the refined Chern-Simons knot invariants, we had to evaluate it at $\lambda= -t^N \sqrt{t/q}$ (this is equivalent to $\lambda = - {\bf q}^N/{\bf t},$ in knot theory variables). But, this is nothing but the index of the $d_N$ differential:
In the present language, $d_N$ has $(S_1, S_2, Q)$ grades $(1/2, 1/2+N, 1)$, so evaluating $q^{S_1} t^{-S_2}\lambda^Q$ at
$\lambda= -t^N \sqrt{t/q}$, it anti-commutes with $d_N$, turning the superpolynomial into the corresponding index.
This also explains how it is that refined Chern-Simons theory, which a priory computes an index, ends up having enough information to compute the Poincare polynomial: provided $N$ is sufficiently large, there are no cancellations in the index; the Poincare polynomial and the index agree. On the other hand, both the index and the Poincare polynomial are polynomials in $\lambda$ or $t^N$, so to knowing them at sufficiently large $N$ suffices to determine them for all $N$.

\newsec{ Conclusion }

To summarize, the Poincare polynomial of the theory of $N$ M5 branes wrapping $M$ in $Y=T^*M$, %
\eqn\ppoly{
{\rm Tr}_{{\cal H}_{BPS}} (-1)^{2S_1} q^{S_1} t^{-S_2}
}
counting BPS states along $\,\IC \times S^1$ inside ${TN \times S^1}$
is expected \wr\ to compute the dimensions of  $SU(N)$ knot homology groups, when one introduces knots  $K$ in the theory (by, for example, adding additional M5 branes wrapping $L_K \times \IC \times S^1$, or ). One can study the theory either directly, as in \wr\ or at large $N$, as originally proposed in \gsv . This is defined for an arbitrary knot $K$ and arbitrary three manifold $M$. However, it is hard to compute in general, as one needs to know the space of BPS states ${\cal H}_{BPS}$.

By contrast, the index of the M5 brane theory,
\eqn\rindexa{
{\rm Tr}_{{\cal H}} (-1)^{2S_1} q^{S_1-S_R} t^{S_R -S_2}
}
is simple to compute when it exists: when $M$ and knots in it admit a semi-free $U(1)$ action, and the theory has an extra grade, $S_R$ associated to it. The index receives contributions only from the space of BPS states, ${\cal H}_{BPS}\subset {\cal H}$. But, as the partition function of the theory, is computable simply by cutting and gluing, from $S$ and $T$ matrices of the refined Chern-Simons theory, given in section $2.$

As explained in \wr\ in the context of the Poincare polynomial \ppoly , and in \ref\kevin{M. Aganagic, K. Schaeffer, "Orientifolds and the Refined Topological String", to appear} in the context of the index \rindexa , both the index and the Poincare polynomial have a natural generalization to other groups of ADE type, by replacing M5 branes by more general ADE type $(2,0)$ theory in the definition of \ppoly\ and \rindexa .  In the case of the index,  the corresponding generalization results in ADE type refined Chern-Simons theories. These were constructed and solved in \kevin, showing that, as expected \AS\ the connection to earlier work of I. Cherednik \ref\cheredniko{I. Cherednik, "Macdonald's Evaluation Conjectures and Difference Fourier Transform",
arXiv:q-alg/9412016} on ADE type Macdonald Polynomials and $SL(2,Z)$ representations emerges.

From the mathematical perspective, we predict that, when $M$ and knots in it admit a semi-free U(1) action, the knot homology groups admit an additional grade,
$$
{\cal H}_{ij} = \oplus_{k} {\cal H}_{ijk}.
$$
This allows one to define a refined index \rindexa , written in terms of knot theory variables abstractly as,
\eqn\rcci{
\sum_{i,j,k} (-1)^k {\bf q}^i {\bf t}^{j+k} {\rm {dim}} {\cal H}_{ijk}.
}
This has more information about knot homology than the Euler characteristic computed by the ordinary Chern-Simons theory of $ADE$ type. The index \rcci\ reduces to the Euler characteristic only upon setting ${\bf t}=-1.$

While the Poincare polynomial of the knot homology theory
\eqn\ppb{
\sum_{i,j, k} {\bf q}^i {\bf t}^{j} {\rm {dim}} {\cal H}_{ijk} =
\sum_{i,j} {\bf q}^i {\bf t}^{j} {\rm {dim}} {\cal H}_{ij}  .
}
has yet more information than the index \rcci , computing it is hard. The index can, by contrast be obtained simply, by cutting and gluing from $S$ and $T$ matrices of Macdonald type, described in the $SU(N)$ case in Sec. 2, and in the general ADE case, in \kevin.

Moreover, the index \rindexa\rcci\ is interesting for another reason. Its robustness implies that string dualities can be used to relate it to other problems in physics and mathematics. For example, large $N$ dualities should relate refined Chern-Simons invariants to partition functions of
five dimensional ${\cal N}=1$ gauge theories in omega background, analogous to \amv\ in the unrefined case. Moreover, the refined topological vertex of \civ\ should be derivable from the refined Chern-Simons theory, similarly to the way the unrefined topological vertex emerged from this \tv . An evidence in this direction is presented in \ref\ki{ A.Iqbal and C.Kozcaz, Refined Hopf Link Revisited, arXiv:1111.0525}, where it was shown that the $S$ matrix of the refined Chern-Simons theory also arizes from the refined topological vertex formalism \civ\giv. This also proves that the Hopf link knot homologies of \giv\ and those arising from the refined Chern-Simons theory are equivalent. Moreover, we expect that it should be possible to define, at least for toric Calabi-Yau, a refined version of mirror symmetry (related to a five dimensional, or $q$-version of AGT correspondence \ref\Awata1{
H. Awata and Y. Yamada, Five-dimensional AGT Relation and the Deformed beta-ensemble, Prog.Theor.Phys.124 (2010) 227-262
}\ref\Awata2{
H.Awata and Y. Yamada, Five-dimensional AGT Conjecture and the Deformed Virasoro Algebra, JHEP 1001 (2010) 125}). Chern-Simons theory is expected to play a role in this.

\newsec{Acknowledgements}

We are grateful to I. Cherednik, E. Gorsky, S. Gukov, A. Okounkov,  K.Schaeffer, C. Vafa and E. Witten for helpful discussions. We are also grateful to the organizers of String-Math Conference 2011 at U. of Pensilvania, 
Mirror Symmetry and Tropical Geometry Conference, Centraro, Italy, Simons Summer Workshop in Mathematics and Physics, 2011 at the Simons Center for Geometry and Physics, and the PCTS Workshop on Exact Methods in Gauge/String Theories, for the opportunity to present our work in a stimulating atmosphere. This research is supported in part by Berkeley Center for Theoretical Physics, by the National Science Foundation (award number 0855653), by the Institute for the Physics and Mathematics of the Universe, by the US Department of Energy under Contract DE-AC02-05CH11231. The work of S.S. is also partly supported by Ministry of Education and Science of the Russian Federation under contract 14.740.11.5194, by RFBR grant 10-01-00536 and by joint grants 09-02-93105-CNRSL, 09-02-91005-ANF.

\appendix{A}{Torus Knot Invariants From Refined Chern-Simons Theory}

In this section we list several examples of the refined Chern-Simons theory knot invariants $Z(S^3, K, \tableau{1})/Z(S^3, \bigcirc, \tableau{1})$, colored by the fundamental representation, and written in terms of
$$
{\bf t } = \sqrt{t} , \ \ \ {\bf q} = -\sqrt{q/t}, \ \ \ {\bf a}^2 = t^{N} \sqrt{t/q},
$$
for various torus knots $K$. To the best of our knowledge, they agree with the corresponding superpolynomials ${\cal P}(K)({\bf a},{\bf q},{\bf t}),$ provided to us by E. Gorsky,
$$Z(S^3, K, \tableau{1})/Z(S^3, \bigcirc, \tableau{1}) = {\cal P}(K)({\bf a},{\bf q},{\bf t}),
$$
We choose to normalize the invariant in such a way, that ${\cal P} = 1 + O({\bf a},{\bf q},{\bf t})$.

\bigskip

$
{\cal P}(K_{3,7}) = 1+{\bf q}^4 {\bf t}^2+{\bf q}^6 {\bf t}^4+{\bf q}^8 {\bf t}^4+{\bf q}^{10} {\bf t}^6+{\bf q}^{12} {\bf t}^6+{\bf q}^{12} {\bf t}^8+{\bf q}^{14} {\bf t}^8+{\bf q}^{16} {\bf t}^8+{\bf q}^{18} {\bf t}^{10}+{\bf q}^{20} {\bf t}^{10}+{\bf q}^{24} {\bf t}^{12}+({\bf q}^2 {\bf t}^3+{\bf q}^4 {\bf t}^5+{\bf q}^6 {\bf t}^5+2 {\bf q}^8 {\bf t}^7+{\bf q}^{10} {\bf t}^7+{\bf q}^{10} {\bf t}^9+2 {\bf q}^{12} {\bf t}^9+{\bf q}^{14} {\bf t}^9+{\bf q}^{14} {\bf t}^{11}+2 {\bf q}^{16} {\bf t}^{11}+{\bf q}^{18} {\bf t}^{11}+{\bf q}^{20} {\bf t}^{13}+{\bf q}^{22} {\bf t}^{13}) {\bf a}^2+({\bf q}^6 {\bf t}^8+{\bf q}^{10} {\bf t}^{10}+{\bf q}^{12} {\bf t}^{12}+{\bf q}^{14} {\bf t}^{12}+{\bf q}^{18} {\bf t}^{14}) {\bf a}^4
$
\bigskip

$
{\cal P}(K_{4,7}) = 1+{\bf q}^{22} {\bf t}^{14}+{\bf q}^4 {\bf t}^2+{\bf q}^6 {\bf t}^4+{\bf q}^8 {\bf t}^4+{\bf q}^8 {\bf t}^6+{\bf q}^{10} {\bf t}^6+{\bf q}^{12} {\bf t}^6+{\bf q}^{30} {\bf t}^{16}+{\bf q}^{28} {\bf t}^{14}+{\bf q}^{22} {\bf t}^{12}+2 {\bf q}^{16} {\bf t}^{10}+2 {\bf q}^{12} {\bf t}^8+{\bf q}^{36} {\bf t}^{18}+{\bf q}^{32} {\bf t}^{16}+{\bf q}^{14} {\bf t}^{10}+{\bf q}^{18} {\bf t}^{12}+2 {\bf q}^{20} {\bf t}^{12}+{\bf q}^{14} {\bf t}^8+{\bf q}^{28} {\bf t}^{16}+{\bf q}^{16} {\bf t}^8+{\bf q}^{18} {\bf t}^{10}+{\bf q}^{20} {\bf t}^{10}+{\bf q}^{26} {\bf t}^{14}+2 {\bf q}^{24} {\bf t}^{14}+{\bf q}^{24} {\bf t}^{12}+({\bf q}^2 {\bf t}^3+{\bf q}^4 {\bf t}^5+{\bf q}^6 {\bf t}^5+{\bf q}^6 {\bf t}^7+2 {\bf q}^8 {\bf t}^7+{\bf q}^{10} {\bf t}^7+3 {\bf q}^{10} {\bf t}^9+2 {\bf q}^{12} {\bf t}^9+2 {\bf q}^{12} {\bf t}^{11}+{\bf q}^{14} {\bf t}^9+4 {\bf q}^{14} {\bf t}^{11}+2 {\bf q}^{16} {\bf t}^{11}+3 {\bf q}^{16} {\bf t}^{13}+{\bf q}^{18} {\bf t}^{11}+4 {\bf q}^{18} {\bf t}^{13}+{\bf q}^{18} {\bf t}^{15}+2 {\bf q}^{20} {\bf t}^{13}+3 {\bf q}^{20} {\bf t}^{15}+{\bf q}^{22} {\bf t}^{13}+4 {\bf q}^{22} {\bf t}^{15}+2 {\bf q}^{24} {\bf t}^{15}+2 {\bf q}^{24} {\bf t}^{17}+{\bf q}^{26} {\bf t}^{15}+3 {\bf q}^{26} {\bf t}^{17}+2 {\bf q}^{28} {\bf t}^{17}+{\bf q}^{30} {\bf t}^{17}+{\bf q}^{30} {\bf t}^{19}+{\bf q}^{32} {\bf t}^{19}+{\bf q}^{34} {\bf t}^{19}) {\bf a}^2+({\bf q}^6 {\bf t}^8+{\bf q}^8 {\bf t}^{10}+{\bf q}^{10} {\bf t}^{10}+{\bf q}^{10} {\bf t}^{12}+2 {\bf q}^{12} {\bf t}^{12}+{\bf q}^{14} {\bf t}^{12}+3 {\bf q}^{14} {\bf t}^{14}+2 {\bf q}^{16} {\bf t}^{14}+{\bf q}^{16} {\bf t}^{16}+{\bf q}^{18} {\bf t}^{14}+3 {\bf q}^{18} {\bf t}^{16}+2 {\bf q}^{20} {\bf t}^{16}+{\bf q}^{20} {\bf t}^{18}+{\bf q}^{22} {\bf t}^{16}+3 {\bf q}^{22} {\bf t}^{18}+2 {\bf q}^{24} {\bf t}^{18}+{\bf q}^{26} {\bf t}^{18}+{\bf q}^{26} {\bf t}^{20}+{\bf q}^{28} {\bf t}^{20}+{\bf q}^{30} {\bf t}^{20}) {\bf a}^4+({\bf q}^{12} {\bf t}^{15}+{\bf q}^{16} {\bf t}^{17}+{\bf q}^{18} {\bf t}^{19}+{\bf q}^{20} {\bf t}^{19}+{\bf q}^{24} {\bf t}^{21}) {\bf a}^6
$
\bigskip

$
{\cal P}(K_{5,8}) = 1+2 {\bf q}^{40} {\bf t}^{22}+2 {\bf q}^{32} {\bf t}^{18}+{\bf q}^{32} {\bf t}^{16}+3 {\bf q}^{38} {\bf t}^{22}+3 {\bf q}^{34} {\bf t}^{20}+2 {\bf q}^{42} {\bf t}^{24}+{\bf q}^{40} {\bf t}^{20}+{\bf q}^{38} {\bf t}^{20}+3 {\bf q}^{36} {\bf t}^{22}+2 {\bf q}^{36} {\bf t}^{20}+{\bf q}^{34} {\bf t}^{18}+{\bf q}^{36} {\bf t}^{18}+{\bf q}^4 {\bf t}^2+{\bf q}^6 {\bf t}^4+{\bf q}^8 {\bf t}^4+{\bf q}^{34} {\bf t}^{22}+{\bf q}^8 {\bf t}^6+{\bf q}^{10} {\bf t}^6+{\bf q}^{12} {\bf t}^6+{\bf q}^{10} {\bf t}^8+{\bf q}^{22} {\bf t}^{12}+2 {\bf q}^{12} {\bf t}^8+2 {\bf q}^{16} {\bf t}^{12}+4 {\bf q}^{24} {\bf t}^{16}+{\bf q}^{22} {\bf t}^{16}+3 {\bf q}^{22} {\bf t}^{14}+4 {\bf q}^{32} {\bf t}^{20}+{\bf q}^{24} {\bf t}^{12}+3 {\bf q}^{20} {\bf t}^{14}+2 {\bf q}^{20} {\bf t}^{12}+3 {\bf q}^{18} {\bf t}^{12}+2 {\bf q}^{14} {\bf t}^{10}+2 {\bf q}^{24} {\bf t}^{14}+3 {\bf q}^{26} {\bf t}^{16}+2 {\bf q}^{26} {\bf t}^{18}+2 {\bf q}^{40} {\bf t}^{24}+{\bf q}^{14} {\bf t}^8+{\bf q}^{42} {\bf t}^{22}+2 {\bf q}^{30} {\bf t}^{20}+3 {\bf q}^{30} {\bf t}^{18}+{\bf q}^{26} {\bf t}^{14}+{\bf q}^{28} {\bf t}^{14}+4 {\bf q}^{28} {\bf t}^{18}+2 {\bf q}^{28} {\bf t}^{16}+{\bf q}^{30} {\bf t}^{16}+{\bf q}^{16} {\bf t}^8+{\bf q}^{18} {\bf t}^{10}+{\bf q}^{20} {\bf t}^{10}+2 {\bf q}^{16} {\bf t}^{10}+{\bf q}^{46} {\bf t}^{24}+{\bf q}^{44} {\bf t}^{22}+2 {\bf q}^{44} {\bf t}^{24}+{\bf q}^{46} {\bf t}^{26}+{\bf q}^{48} {\bf t}^{24}+{\bf q}^{48} {\bf t}^{26}+{\bf q}^{50} {\bf t}^{26}+{\bf q}^{52} {\bf t}^{26}+{\bf q}^{56} {\bf t}^{28}+({\bf q}^2 {\bf t}^3+{\bf q}^4 {\bf t}^5+{\bf q}^6 {\bf t}^5+{\bf q}^6 {\bf t}^7+2 {\bf q}^8 {\bf t}^7+{\bf q}^8 {\bf t}^9+{\bf q}^{10} {\bf t}^7+3 {\bf q}^{10} {\bf t}^9+2 {\bf q}^{12} {\bf t}^9+4 {\bf q}^{12} {\bf t}^{11}+{\bf q}^{14} {\bf t}^9+4 {\bf q}^{14} {\bf t}^{11}+3 {\bf q}^{14} {\bf t}^{13}+2 {\bf q}^{16} {\bf t}^{11}+6 {\bf q}^{16} {\bf t}^{13}+{\bf q}^{18} {\bf t}^{11}+{\bf q}^{16} {\bf t}^{15}+4 {\bf q}^{18} {\bf t}^{13}+7 {\bf q}^{18} {\bf t}^{15}+2 {\bf q}^{20} {\bf t}^{13}+7 {\bf q}^{20} {\bf t}^{15}+{\bf q}^{22} {\bf t}^{13}+4 {\bf q}^{20} {\bf t}^{17}+4 {\bf q}^{22} {\bf t}^{15}+9 {\bf q}^{22} {\bf t}^{17}+2 {\bf q}^{24} {\bf t}^{15}+{\bf q}^{22} {\bf t}^{19}+7 {\bf q}^{24} {\bf t}^{17}+{\bf q}^{26} {\bf t}^{15}+7 {\bf q}^{24} {\bf t}^{19}+4 {\bf q}^{26} {\bf t}^{17}+10 {\bf q}^{26} {\bf t}^{19}+2 {\bf q}^{28} {\bf t}^{17}+2 {\bf q}^{26} {\bf t}^{21}+7 {\bf q}^{28} {\bf t}^{19}+{\bf q}^{30} {\bf t}^{17}+8 {\bf q}^{28} {\bf t}^{21}+4 {\bf q}^{30} {\bf t}^{19}+10 {\bf q}^{30} {\bf t}^{21}+2 {\bf q}^{32} {\bf t}^{19}+2 {\bf q}^{30} {\bf t}^{23}+7 {\bf q}^{32} {\bf t}^{21}+{\bf q}^{34} {\bf t}^{19}+7 {\bf q}^{32} {\bf t}^{23}+4 {\bf q}^{34} {\bf t}^{21}+9 {\bf q}^{34} {\bf t}^{23}+2 {\bf q}^{36} {\bf t}^{21}+{\bf q}^{34} {\bf t}^{25}+7 {\bf q}^{36} {\bf t}^{23}+{\bf q}^{38} {\bf t}^{21}+4 {\bf q}^{36} {\bf t}^{25}+4 {\bf q}^{38} {\bf t}^{23}+7 {\bf q}^{38} {\bf t}^{25}+2 {\bf q}^{40} {\bf t}^{23}+6 {\bf q}^{40} {\bf t}^{25}+{\bf q}^{42} {\bf t}^{23}+{\bf q}^{40} {\bf t}^{27}+4 {\bf q}^{42} {\bf t}^{25}+3 {\bf q}^{42} {\bf t}^{27}+2 {\bf q}^{44} {\bf t}^{25}+4 {\bf q}^{44} {\bf t}^{27}+{\bf q}^{46} {\bf t}^{25}+3 {\bf q}^{46} {\bf t}^{27}+2 {\bf q}^{48} {\bf t}^{27}+{\bf q}^{48} {\bf t}^{29}+{\bf q}^{50} {\bf t}^{27}+{\bf q}^{50} {\bf t}^{29}+{\bf q}^{52} {\bf t}^{29}+{\bf q}^{54} {\bf t}^{29}) {\bf a}^2+({\bf q}^6 {\bf t}^8+{\bf q}^8 {\bf t}^{10}+{\bf q}^{10} {\bf t}^{10}+2 {\bf q}^{10} {\bf t}^{12}+2 {\bf q}^{12} {\bf t}^{12}+{\bf q}^{12} {\bf t}^{14}+{\bf q}^{14} {\bf t}^{12}+4 {\bf q}^{14} {\bf t}^{14}+{\bf q}^{14} {\bf t}^{16}+2 {\bf q}^{16} {\bf t}^{14}+5 {\bf q}^{16} {\bf t}^{16}+{\bf q}^{18} {\bf t}^{14}+5 {\bf q}^{18} {\bf t}^{16}+4 {\bf q}^{18} {\bf t}^{18}+2 {\bf q}^{20} {\bf t}^{16}+7 {\bf q}^{20} {\bf t}^{18}+{\bf q}^{22} {\bf t}^{16}+2 {\bf q}^{20} {\bf t}^{20}+5 {\bf q}^{22} {\bf t}^{18}+8 {\bf q}^{22} {\bf t}^{20}+2 {\bf q}^{24} {\bf t}^{18}+8 {\bf q}^{24} {\bf t}^{20}+{\bf q}^{26} {\bf t}^{18}+4 {\bf q}^{24} {\bf t}^{22}+5 {\bf q}^{26} {\bf t}^{20}+10 {\bf q}^{26} {\bf t}^{22}+2 {\bf q}^{28} {\bf t}^{20}+{\bf q}^{26} {\bf t}^{24}+8 {\bf q}^{28} {\bf t}^{22}+{\bf q}^{30} {\bf t}^{20}+5 {\bf q}^{28} {\bf t}^{24}+5 {\bf q}^{30} {\bf t}^{22}+10 {\bf q}^{30} {\bf t}^{24}+2 {\bf q}^{32} {\bf t}^{22}+{\bf q}^{30} {\bf t}^{26}+8 {\bf q}^{32} {\bf t}^{24}+{\bf q}^{34} {\bf t}^{22}+4 {\bf q}^{32} {\bf t}^{26}+5 {\bf q}^{34} {\bf t}^{24}+8 {\bf q}^{34} {\bf t}^{26}+2 {\bf q}^{36} {\bf t}^{24}+7 {\bf q}^{36} {\bf t}^{26}+{\bf q}^{38} {\bf t}^{24}+2 {\bf q}^{36} {\bf t}^{28}+5 {\bf q}^{38} {\bf t}^{26}+4 {\bf q}^{38} {\bf t}^{28}+2 {\bf q}^{40} {\bf t}^{26}+5 {\bf q}^{40} {\bf t}^{28}+{\bf q}^{42} {\bf t}^{26}+4 {\bf q}^{42} {\bf t}^{28}+{\bf q}^{42} {\bf t}^{30}+2 {\bf q}^{44} {\bf t}^{28}+{\bf q}^{44} {\bf t}^{30}+{\bf q}^{46} {\bf t}^{28}+2 {\bf q}^{46} {\bf t}^{30}+{\bf q}^{48} {\bf t}^{30}+{\bf q}^{50} {\bf t}^{30}) {\bf a}^4+({\bf q}^{12} {\bf t}^{15}+{\bf q}^{14} {\bf t}^{17}+{\bf q}^{16} {\bf t}^{17}+{\bf q}^{16} {\bf t}^{19}+2 {\bf q}^{18} {\bf t}^{19}+{\bf q}^{18} {\bf t}^{21}+{\bf q}^{20} {\bf t}^{19}+3 {\bf q}^{20} {\bf t}^{21}+2 {\bf q}^{22} {\bf t}^{21}+3 {\bf q}^{22} {\bf t}^{23}+{\bf q}^{24} {\bf t}^{21}+4 {\bf q}^{24} {\bf t}^{23}+{\bf q}^{24} {\bf t}^{25}+2 {\bf q}^{26} {\bf t}^{23}+4 {\bf q}^{26} {\bf t}^{25}+{\bf q}^{28} {\bf t}^{23}+4 {\bf q}^{28} {\bf t}^{25}+2 {\bf q}^{28} {\bf t}^{27}+2 {\bf q}^{30} {\bf t}^{25}+4 {\bf q}^{30} {\bf t}^{27}+{\bf q}^{32} {\bf t}^{25}+4 {\bf q}^{32} {\bf t}^{27}+{\bf q}^{32} {\bf t}^{29}+2 {\bf q}^{34} {\bf t}^{27}+3 {\bf q}^{34} {\bf t}^{29}+{\bf q}^{36} {\bf t}^{27}+3 {\bf q}^{36} {\bf t}^{29}+2 {\bf q}^{38} {\bf t}^{29}+{\bf q}^{38} {\bf t}^{31}+{\bf q}^{40} {\bf t}^{29}+{\bf q}^{40} {\bf t}^{31}+{\bf q}^{42} {\bf t}^{31}+{\bf q}^{44} {\bf t}^{31}) {\bf a}^6+({\bf q}^{20} {\bf t}^{24}+{\bf q}^{24} {\bf t}^{26}+{\bf q}^{26} {\bf t}^{28}+{\bf q}^{28} {\bf t}^{28}+{\bf q}^{30} {\bf t}^{30}+{\bf q}^{32} {\bf t}^{30}+{\bf q}^{36} {\bf t}^{32}) {\bf a}^8
$
\bigskip

$
{\cal P}(K_{7,10}) = 1+2 {\bf q}^{40} {\bf t}^{22}+11 {\bf q}^{38} {\bf t}^{28}+2 {\bf q}^{32} {\bf t}^{18}+{\bf q}^{32} {\bf t}^{16}+3 {\bf q}^{38} {\bf t}^{22}+3 {\bf q}^{34} {\bf t}^{20}+3 {\bf q}^{42} {\bf t}^{24}+15 {\bf q}^{44} {\bf t}^{30}+{\bf q}^{40} {\bf t}^{20}+{\bf q}^{38} {\bf t}^{20}+10 {\bf q}^{44} {\bf t}^{28}+5 {\bf q}^{36} {\bf t}^{22}+2 {\bf q}^{36} {\bf t}^{20}+11 {\bf q}^{44} {\bf t}^{32}+12 {\bf q}^{42} {\bf t}^{28}+10 {\bf q}^{36} {\bf t}^{24}+{\bf q}^{34} {\bf t}^{18}+{\bf q}^{36} {\bf t}^{18}+7 {\bf q}^{40} {\bf t}^{30}+{\bf q}^4 {\bf t}^2+{\bf q}^6 {\bf t}^4+{\bf q}^8 {\bf t}^4+10 {\bf q}^{34} {\bf t}^{24}+7 {\bf q}^{34} {\bf t}^{22}+14 {\bf q}^{42} {\bf t}^{30}+10 {\bf q}^{40} {\bf t}^{26}+7 {\bf q}^{42} {\bf t}^{26}+3 {\bf q}^{36} {\bf t}^{28}+{\bf q}^8 {\bf t}^6+{\bf q}^{10} {\bf t}^6+{\bf q}^{12} {\bf t}^6+{\bf q}^{10} {\bf t}^8+{\bf q}^{22} {\bf t}^{12}+2 {\bf q}^{12} {\bf t}^8+2 {\bf q}^{42} {\bf t}^{32}+{\bf q}^{12} {\bf t}^{10}+3 {\bf q}^{18} {\bf t}^{14}+3 {\bf q}^{16} {\bf t}^{12}+5 {\bf q}^{24} {\bf t}^{16}+5 {\bf q}^{22} {\bf t}^{16}+3 {\bf q}^{22} {\bf t}^{14}+5 {\bf q}^{26} {\bf t}^{20}+5 {\bf q}^{32} {\bf t}^{20}+4 {\bf q}^{28} {\bf t}^{22}+{\bf q}^{24} {\bf t}^{12}+4 {\bf q}^{20} {\bf t}^{14}+3 {\bf q}^{20} {\bf t}^{16}+2 {\bf q}^{20} {\bf t}^{12}+{\bf q}^{24} {\bf t}^{20}+6 {\bf q}^{24} {\bf t}^{18}+{\bf q}^{22} {\bf t}^{18}+3 {\bf q}^{18} {\bf t}^{12}+8 {\bf q}^{28} {\bf t}^{20}+2 {\bf q}^{14} {\bf t}^{10}+2 {\bf q}^{24} {\bf t}^{14}+3 {\bf q}^{26} {\bf t}^{16}+6 {\bf q}^{26} {\bf t}^{18}+5 {\bf q}^{40} {\bf t}^{24}+7 {\bf q}^{38} {\bf t}^{24}+11 {\bf q}^{38} {\bf t}^{26}+4 {\bf q}^{46} {\bf t}^{34}+{\bf q}^{14} {\bf t}^8+14 {\bf q}^{40} {\bf t}^{28}+12 {\bf q}^{36} {\bf t}^{26}+{\bf q}^{42} {\bf t}^{22}+6 {\bf q}^{34} {\bf t}^{26}+7 {\bf q}^{30} {\bf t}^{20}+3 {\bf q}^{30} {\bf t}^{18}+8 {\bf q}^{30} {\bf t}^{22}+2 {\bf q}^{30} {\bf t}^{24}+{\bf q}^{26} {\bf t}^{14}+{\bf q}^{28} {\bf t}^{14}+9 {\bf q}^{32} {\bf t}^{22}+9 {\bf q}^{32} {\bf t}^{24}+5 {\bf q}^{28} {\bf t}^{18}+2 {\bf q}^{28} {\bf t}^{16}+{\bf q}^{30} {\bf t}^{16}+{\bf q}^{16} {\bf t}^8+{\bf q}^{18} {\bf t}^{10}+{\bf q}^{20} {\bf t}^{10}+2 {\bf q}^{16} {\bf t}^{10}+{\bf q}^{14} {\bf t}^{12}+{\bf q}^{56} {\bf t}^{40}+{\bf q}^{46} {\bf t}^{24}+{\bf q}^{98} {\bf t}^{52}+{\bf q}^{84} {\bf t}^{50}+{\bf q}^{70} {\bf t}^{36}+{\bf q}^{72} {\bf t}^{36}+{\bf q}^{68} {\bf t}^{34}+{\bf q}^{82} {\bf t}^{42}+{\bf q}^{96} {\bf t}^{52}+{\bf q}^{84} {\bf t}^{42}+{\bf q}^{94} {\bf t}^{52}+{\bf q}^{92} {\bf t}^{46}+{\bf q}^{94} {\bf t}^{48}+{\bf q}^{78} {\bf t}^{40}+{\bf q}^{90} {\bf t}^{46}+{\bf q}^{98} {\bf t}^{50}+{\bf q}^{102} {\bf t}^{52}+{\bf q}^{96} {\bf t}^{48}+{\bf q}^{104} {\bf t}^{52}+{\bf q}^{100} {\bf t}^{50}+{\bf q}^{88} {\bf t}^{44}+{\bf q}^{100} {\bf t}^{52}+{\bf q}^{86} {\bf t}^{44}+{\bf q}^{108} {\bf t}^{54}+{\bf q}^{86} {\bf t}^{50}+9 {\bf q}^{76} {\bf t}^{46}+14 {\bf q}^{66} {\bf t}^{42}+8 {\bf q}^{78} {\bf t}^{46}+5 {\bf q}^{86} {\bf t}^{48}+7 {\bf q}^{68} {\bf t}^{44}+11 {\bf q}^{64} {\bf t}^{42}+2 {\bf q}^{66} {\bf t}^{44}+{\bf q}^{44} {\bf t}^{22}+5 {\bf q}^{44} {\bf t}^{26}+7 {\bf q}^{46} {\bf t}^{28}+12 {\bf q}^{46} {\bf t}^{30}+{\bf q}^{48} {\bf t}^{36}+2 {\bf q}^{44} {\bf t}^{24}+16 {\bf q}^{46} {\bf t}^{32}+3 {\bf q}^{46} {\bf t}^{26}+{\bf q}^{74} {\bf t}^{38}+{\bf q}^{80} {\bf t}^{40}+{\bf q}^{76} {\bf t}^{38}+{\bf q}^{62} {\bf t}^{32}+{\bf q}^{66} {\bf t}^{34}+2 {\bf q}^{72} {\bf t}^{38}+10 {\bf q}^{56} {\bf t}^{34}+7 {\bf q}^{58} {\bf t}^{34}+5 {\bf q}^{64} {\bf t}^{36}+3 {\bf q}^{62} {\bf t}^{34}+2 {\bf q}^{64} {\bf t}^{34}+3 {\bf q}^{66} {\bf t}^{36}+5 {\bf q}^{56} {\bf t}^{32}+5 {\bf q}^{60} {\bf t}^{34}+2 {\bf q}^{68} {\bf t}^{36}+{\bf q}^{64} {\bf t}^{32}+7 {\bf q}^{50} {\bf t}^{30}+16 {\bf q}^{60} {\bf t}^{38}+10 {\bf q}^{64} {\bf t}^{38}+4 {\bf q}^{80} {\bf t}^{48}+9 {\bf q}^{76} {\bf t}^{44}+12 {\bf q}^{72} {\bf t}^{44}+10 {\bf q}^{72} {\bf t}^{42}+7 {\bf q}^{78} {\bf t}^{44}+10 {\bf q}^{68} {\bf t}^{40}+7 {\bf q}^{74} {\bf t}^{42}+11 {\bf q}^{70} {\bf t}^{42}+5 {\bf q}^{72} {\bf t}^{40}+5 {\bf q}^{68} {\bf t}^{38}+3 {\bf q}^{70} {\bf t}^{38}+6 {\bf q}^{82} {\bf t}^{46}+2 {\bf q}^{78} {\bf t}^{48}+16 {\bf q}^{62} {\bf t}^{40}+3 {\bf q}^{72} {\bf t}^{46}+11 {\bf q}^{70} {\bf t}^{44}+12 {\bf q}^{66} {\bf t}^{40}+3 {\bf q}^{92} {\bf t}^{50}+5 {\bf q}^{82} {\bf t}^{48}+15 {\bf q}^{64} {\bf t}^{40}+10 {\bf q}^{74} {\bf t}^{44}+3 {\bf q}^{86} {\bf t}^{46}+3 {\bf q}^{78} {\bf t}^{42}+2 {\bf q}^{76} {\bf t}^{40}+4 {\bf q}^{88} {\bf t}^{48}+6 {\bf q}^{84} {\bf t}^{48}+3 {\bf q}^{74} {\bf t}^{40}+6 {\bf q}^{74} {\bf t}^{46}+5 {\bf q}^{76} {\bf t}^{42}+5 {\bf q}^{80} {\bf t}^{44}+5 {\bf q}^{84} {\bf t}^{46}+2 {\bf q}^{88} {\bf t}^{46}+2 {\bf q}^{96} {\bf t}^{50}+2 {\bf q}^{92} {\bf t}^{48}+8 {\bf q}^{80} {\bf t}^{46}+2 {\bf q}^{80} {\bf t}^{42}+3 {\bf q}^{82} {\bf t}^{44}+2 {\bf q}^{84} {\bf t}^{44}+3 {\bf q}^{90} {\bf t}^{50}+3 {\bf q}^{90} {\bf t}^{48}+2 {\bf q}^{94} {\bf t}^{50}+12 {\bf q}^{62} {\bf t}^{38}+7 {\bf q}^{70} {\bf t}^{40}+2 {\bf q}^{60} {\bf t}^{32}+7 {\bf q}^{62} {\bf t}^{36}+3 {\bf q}^{58} {\bf t}^{32}+10 {\bf q}^{60} {\bf t}^{36}+7 {\bf q}^{66} {\bf t}^{38}+12 {\bf q}^{58} {\bf t}^{36}+3 {\bf q}^{88} {\bf t}^{50}+14 {\bf q}^{68} {\bf t}^{42}+5 {\bf q}^{48} {\bf t}^{28}+10 {\bf q}^{48} {\bf t}^{30}+12 {\bf q}^{50} {\bf t}^{32}+{\bf q}^{48} {\bf t}^{24}+16 {\bf q}^{48} {\bf t}^{32}+2 {\bf q}^{48} {\bf t}^{26}+17 {\bf q}^{50} {\bf t}^{34}+14 {\bf q}^{48} {\bf t}^{34}+{\bf q}^{50} {\bf t}^{26}+6 {\bf q}^{50} {\bf t}^{36}+{\bf q}^{52} {\bf t}^{38}+17 {\bf q}^{58} {\bf t}^{38}+6 {\bf q}^{58} {\bf t}^{40}+3 {\bf q}^{50} {\bf t}^{28}+12 {\bf q}^{54} {\bf t}^{34}+{\bf q}^{52} {\bf t}^{26}+3 {\bf q}^{54} {\bf t}^{30}+5 {\bf q}^{52} {\bf t}^{30}+{\bf q}^{54} {\bf t}^{28}+{\bf q}^{56} {\bf t}^{28}+16 {\bf q}^{56} {\bf t}^{36}+{\bf q}^{60} {\bf t}^{42}+{\bf q}^{58} {\bf t}^{30}+2 {\bf q}^{56} {\bf t}^{30}+7 {\bf q}^{54} {\bf t}^{32}+{\bf q}^{60} {\bf t}^{30}+14 {\bf q}^{60} {\bf t}^{40}+4 {\bf q}^{62} {\bf t}^{42}+7 {\bf q}^{54} {\bf t}^{38}+15 {\bf q}^{52} {\bf t}^{36}+18 {\bf q}^{54} {\bf t}^{36}+16 {\bf q}^{52} {\bf t}^{34}+15 {\bf q}^{56} {\bf t}^{38}+10 {\bf q}^{52} {\bf t}^{32}+2 {\bf q}^{52} {\bf t}^{28}+({\bf q}^2 {\bf t}^3+{\bf q}^4 {\bf t}^5+{\bf q}^6 {\bf t}^5+{\bf q}^6 {\bf t}^7+2 {\bf q}^8 {\bf t}^7+{\bf q}^8 {\bf t}^9+{\bf q}^{10} {\bf t}^7+3 {\bf q}^{10} {\bf t}^9+{\bf q}^{10} {\bf t}^{11}+2 {\bf q}^{12} {\bf t}^9+4 {\bf q}^{12} {\bf t}^{11}+{\bf q}^{14} {\bf t}^9+{\bf q}^{12} {\bf t}^{13}+4 {\bf q}^{14} {\bf t}^{11}+5 {\bf q}^{14} {\bf t}^{13}+2 {\bf q}^{16} {\bf t}^{11}+6 {\bf q}^{16} {\bf t}^{13}+{\bf q}^{18} {\bf t}^{11}+6 {\bf q}^{16} {\bf t}^{15}+4 {\bf q}^{18} {\bf t}^{13}+9 {\bf q}^{18} {\bf t}^{15}+2 {\bf q}^{20} {\bf t}^{13}+5 {\bf q}^{18} {\bf t}^{17}+7 {\bf q}^{20} {\bf t}^{15}+{\bf q}^{22} {\bf t}^{13}+12 {\bf q}^{20} {\bf t}^{17}+4 {\bf q}^{22} {\bf t}^{15}+3 {\bf q}^{20} {\bf t}^{19}+11 {\bf q}^{22} {\bf t}^{17}+2 {\bf q}^{24} {\bf t}^{15}+14 {\bf q}^{22} {\bf t}^{19}+7 {\bf q}^{24} {\bf t}^{17}+{\bf q}^{26} {\bf t}^{15}+2 {\bf q}^{22} {\bf t}^{21}+16 {\bf q}^{24} {\bf t}^{19}+4 {\bf q}^{26} {\bf t}^{17}+13 {\bf q}^{24} {\bf t}^{21}+12 {\bf q}^{26} {\bf t}^{19}+2 {\bf q}^{28} {\bf t}^{17}+{\bf q}^{24} {\bf t}^{23}+21 {\bf q}^{26} {\bf t}^{21}+7 {\bf q}^{28} {\bf t}^{19}+{\bf q}^{30} {\bf t}^{17}+11 {\bf q}^{26} {\bf t}^{23}+18 {\bf q}^{28} {\bf t}^{21}+4 {\bf q}^{30} {\bf t}^{19}+24 {\bf q}^{28} {\bf t}^{23}+12 {\bf q}^{30} {\bf t}^{21}+2 {\bf q}^{32} {\bf t}^{19}+7 {\bf q}^{28} {\bf t}^{25}+25 {\bf q}^{30} {\bf t}^{23}+7 {\bf q}^{32} {\bf t}^{21}+{\bf q}^{34} {\bf t}^{19}+25 {\bf q}^{30} {\bf t}^{25}+19 {\bf q}^{32} {\bf t}^{23}+4 {\bf q}^{34} {\bf t}^{21}+3 {\bf q}^{30} {\bf t}^{27}+31 {\bf q}^{32} {\bf t}^{25}+12 {\bf q}^{34} {\bf t}^{23}+2 {\bf q}^{36} {\bf t}^{21}+21 {\bf q}^{32} {\bf t}^{27}+27 {\bf q}^{34} {\bf t}^{25}+7 {\bf q}^{36} {\bf t}^{23}+{\bf q}^{38} {\bf t}^{21}+{\bf q}^{32} {\bf t}^{29}+37 {\bf q}^{34} {\bf t}^{27}+19 {\bf q}^{36} {\bf t}^{25}+4 {\bf q}^{38} {\bf t}^{23}+13 {\bf q}^{34} {\bf t}^{29}+35 {\bf q}^{36} {\bf t}^{27}+12 {\bf q}^{38} {\bf t}^{25}+2 {\bf q}^{40} {\bf t}^{23}+37 {\bf q}^{36} {\bf t}^{29}+28 {\bf q}^{38} {\bf t}^{27}+7 {\bf q}^{40} {\bf t}^{25}+{\bf q}^{42} {\bf t}^{23}+6 {\bf q}^{36} {\bf t}^{31}+44 {\bf q}^{38} {\bf t}^{29}+19 {\bf q}^{40} {\bf t}^{27}+4 {\bf q}^{42} {\bf t}^{25}+29 {\bf q}^{38} {\bf t}^{31}+37 {\bf q}^{40} {\bf t}^{29}+12 {\bf q}^{42} {\bf t}^{27}+2 {\bf q}^{44} {\bf t}^{25}+{\bf q}^{38} {\bf t}^{33}+49 {\bf q}^{40} {\bf t}^{31}+28 {\bf q}^{42} {\bf t}^{29}+7 {\bf q}^{44} {\bf t}^{27}+{\bf q}^{46} {\bf t}^{25}+16 {\bf q}^{40} {\bf t}^{33}+48 {\bf q}^{42} {\bf t}^{31}+19 {\bf q}^{44} {\bf t}^{29}+4 {\bf q}^{46} {\bf t}^{27}+44 {\bf q}^{42} {\bf t}^{33}+38 {\bf q}^{44} {\bf t}^{31}+12 {\bf q}^{46} {\bf t}^{29}+2 {\bf q}^{48} {\bf t}^{27}+6 {\bf q}^{42} {\bf t}^{35}+56 {\bf q}^{44} {\bf t}^{33}+28 {\bf q}^{46} {\bf t}^{31}+7 {\bf q}^{48} {\bf t}^{29}+{\bf q}^{50} {\bf t}^{27}+29 {\bf q}^{44} {\bf t}^{35}+50 {\bf q}^{46} {\bf t}^{33}+19 {\bf q}^{48} {\bf t}^{31}+4 {\bf q}^{50} {\bf t}^{29}+{\bf q}^{44} {\bf t}^{37}+55 {\bf q}^{46} {\bf t}^{35}+38 {\bf q}^{48} {\bf t}^{33}+12 {\bf q}^{50} {\bf t}^{31}+2 {\bf q}^{52} {\bf t}^{29}+12 {\bf q}^{46} {\bf t}^{37}+60 {\bf q}^{48} {\bf t}^{35}+28 {\bf q}^{50} {\bf t}^{33}+7 {\bf q}^{52} {\bf t}^{31}+{\bf q}^{54} {\bf t}^{29}+39 {\bf q}^{48} {\bf t}^{37}+51 {\bf q}^{50} {\bf t}^{35}+19 {\bf q}^{52} {\bf t}^{33}+4 {\bf q}^{54} {\bf t}^{31}+3 {\bf q}^{48} {\bf t}^{39}+61 {\bf q}^{50} {\bf t}^{37}+38 {\bf q}^{52} {\bf t}^{35}+12 {\bf q}^{54} {\bf t}^{33}+2 {\bf q}^{56} {\bf t}^{31}+18 {\bf q}^{50} {\bf t}^{39}+62 {\bf q}^{52} {\bf t}^{37}+28 {\bf q}^{54} {\bf t}^{35}+7 {\bf q}^{56} {\bf t}^{33}+{\bf q}^{58} {\bf t}^{31}+45 {\bf q}^{52} {\bf t}^{39}+51 {\bf q}^{54} {\bf t}^{37}+19 {\bf q}^{56} {\bf t}^{35}+4 {\bf q}^{58} {\bf t}^{33}+4 {\bf q}^{52} {\bf t}^{41}+63 {\bf q}^{54} {\bf t}^{39}+38 {\bf q}^{56} {\bf t}^{37}+12 {\bf q}^{58} {\bf t}^{35}+2 {\bf q}^{60} {\bf t}^{33}+19 {\bf q}^{54} {\bf t}^{41}+62 {\bf q}^{56} {\bf t}^{39}+28 {\bf q}^{58} {\bf t}^{37}+7 {\bf q}^{60} {\bf t}^{35}+{\bf q}^{62} {\bf t}^{33}+{\bf q}^{54} {\bf t}^{43}+45 {\bf q}^{56} {\bf t}^{41}+51 {\bf q}^{58} {\bf t}^{39}+19 {\bf q}^{60} {\bf t}^{37}+4 {\bf q}^{62} {\bf t}^{35}+4 {\bf q}^{56} {\bf t}^{43}+61 {\bf q}^{58} {\bf t}^{41}+38 {\bf q}^{60} {\bf t}^{39}+12 {\bf q}^{62} {\bf t}^{37}+2 {\bf q}^{64} {\bf t}^{35}+18 {\bf q}^{58} {\bf t}^{43}+60 {\bf q}^{60} {\bf t}^{41}+28 {\bf q}^{62} {\bf t}^{39}+7 {\bf q}^{64} {\bf t}^{37}+{\bf q}^{66} {\bf t}^{35}+39 {\bf q}^{60} {\bf t}^{43}+50 {\bf q}^{62} {\bf t}^{41}+19 {\bf q}^{64} {\bf t}^{39}+4 {\bf q}^{66} {\bf t}^{37}+3 {\bf q}^{60} {\bf t}^{45}+55 {\bf q}^{62} {\bf t}^{43}+38 {\bf q}^{64} {\bf t}^{41}+12 {\bf q}^{66} {\bf t}^{39}+2 {\bf q}^{68} {\bf t}^{37}+12 {\bf q}^{62} {\bf t}^{45}+56 {\bf q}^{64} {\bf t}^{43}+28 {\bf q}^{66} {\bf t}^{41}+7 {\bf q}^{68} {\bf t}^{39}+{\bf q}^{70} {\bf t}^{37}+29 {\bf q}^{64} {\bf t}^{45}+48 {\bf q}^{66} {\bf t}^{43}+19 {\bf q}^{68} {\bf t}^{41}+4 {\bf q}^{70} {\bf t}^{39}+{\bf q}^{64} {\bf t}^{47}+44 {\bf q}^{66} {\bf t}^{45}+37 {\bf q}^{68} {\bf t}^{43}+12 {\bf q}^{70} {\bf t}^{41}+2 {\bf q}^{72} {\bf t}^{39}+6 {\bf q}^{66} {\bf t}^{47}+49 {\bf q}^{68} {\bf t}^{45}+28 {\bf q}^{70} {\bf t}^{43}+7 {\bf q}^{72} {\bf t}^{41}+{\bf q}^{74} {\bf t}^{39}+16 {\bf q}^{68} {\bf t}^{47}+44 {\bf q}^{70} {\bf t}^{45}+19 {\bf q}^{72} {\bf t}^{43}+4 {\bf q}^{74} {\bf t}^{41}+29 {\bf q}^{70} {\bf t}^{47}+35 {\bf q}^{72} {\bf t}^{45}+12 {\bf q}^{74} {\bf t}^{43}+2 {\bf q}^{76} {\bf t}^{41}+{\bf q}^{70} {\bf t}^{49}+37 {\bf q}^{72} {\bf t}^{47}+27 {\bf q}^{74} {\bf t}^{45}+7 {\bf q}^{76} {\bf t}^{43}+{\bf q}^{78} {\bf t}^{41}+6 {\bf q}^{72} {\bf t}^{49}+37 {\bf q}^{74} {\bf t}^{47}+19 {\bf q}^{76} {\bf t}^{45}+4 {\bf q}^{78} {\bf t}^{43}+13 {\bf q}^{74} {\bf t}^{49}+31 {\bf q}^{76} {\bf t}^{47}+12 {\bf q}^{78} {\bf t}^{45}+2 {\bf q}^{80} {\bf t}^{43}+21 {\bf q}^{76} {\bf t}^{49}+25 {\bf q}^{78} {\bf t}^{47}+7 {\bf q}^{80} {\bf t}^{45}+{\bf q}^{82} {\bf t}^{43}+{\bf q}^{76} {\bf t}^{51}+25 {\bf q}^{78} {\bf t}^{49}+18 {\bf q}^{80} {\bf t}^{47}+4 {\bf q}^{82} {\bf t}^{45}+3 {\bf q}^{78} {\bf t}^{51}+24 {\bf q}^{80} {\bf t}^{49}+12 {\bf q}^{82} {\bf t}^{47}+2 {\bf q}^{84} {\bf t}^{45}+7 {\bf q}^{80} {\bf t}^{51}+21 {\bf q}^{82} {\bf t}^{49}+7 {\bf q}^{84} {\bf t}^{47}+{\bf q}^{86} {\bf t}^{45}+11 {\bf q}^{82} {\bf t}^{51}+16 {\bf q}^{84} {\bf t}^{49}+4 {\bf q}^{86} {\bf t}^{47}+13 {\bf q}^{84} {\bf t}^{51}+11 {\bf q}^{86} {\bf t}^{49}+2 {\bf q}^{88} {\bf t}^{47}+{\bf q}^{84} {\bf t}^{53}+14 {\bf q}^{86} {\bf t}^{51}+7 {\bf q}^{88} {\bf t}^{49}+{\bf q}^{90} {\bf t}^{47}+2 {\bf q}^{86} {\bf t}^{53}+12 {\bf q}^{88} {\bf t}^{51}+4 {\bf q}^{90} {\bf t}^{49}+3 {\bf q}^{88} {\bf t}^{53}+9 {\bf q}^{90} {\bf t}^{51}+2 {\bf q}^{92} {\bf t}^{49}+5 {\bf q}^{90} {\bf t}^{53}+6 {\bf q}^{92} {\bf t}^{51}+{\bf q}^{94} {\bf t}^{49}+6 {\bf q}^{92} {\bf t}^{53}+4 {\bf q}^{94} {\bf t}^{51}+5 {\bf q}^{94} {\bf t}^{53}+2 {\bf q}^{96} {\bf t}^{51}+4 {\bf q}^{96} {\bf t}^{53}+{\bf q}^{98} {\bf t}^{51}+{\bf q}^{96} {\bf t}^{55}+3 {\bf q}^{98} {\bf t}^{53}+{\bf q}^{98} {\bf t}^{55}+2 {\bf q}^{100} {\bf t}^{53}+{\bf q}^{100} {\bf t}^{55}+{\bf q}^{102} {\bf t}^{53}+{\bf q}^{102} {\bf t}^{55}+{\bf q}^{104} {\bf t}^{55}+{\bf q}^{106} {\bf t}^{55}) {\bf a}^2+({\bf q}^6 {\bf t}^8+{\bf q}^8 {\bf t}^{10}+{\bf q}^{10} {\bf t}^{10}+2 {\bf q}^{10} {\bf t}^{12}+2 {\bf q}^{12} {\bf t}^{12}+2 {\bf q}^{12} {\bf t}^{14}+{\bf q}^{14} {\bf t}^{12}+4 {\bf q}^{14} {\bf t}^{14}+3 {\bf q}^{14} {\bf t}^{16}+2 {\bf q}^{16} {\bf t}^{14}+6 {\bf q}^{16} {\bf t}^{16}+{\bf q}^{18} {\bf t}^{14}+2 {\bf q}^{16} {\bf t}^{18}+5 {\bf q}^{18} {\bf t}^{16}+9 {\bf q}^{18} {\bf t}^{18}+2 {\bf q}^{20} {\bf t}^{16}+2 {\bf q}^{18} {\bf t}^{20}+8 {\bf q}^{20} {\bf t}^{18}+{\bf q}^{22} {\bf t}^{16}+11 {\bf q}^{20} {\bf t}^{20}+5 {\bf q}^{22} {\bf t}^{18}+{\bf q}^{20} {\bf t}^{22}+14 {\bf q}^{22} {\bf t}^{20}+2 {\bf q}^{24} {\bf t}^{18}+11 {\bf q}^{22} {\bf t}^{22}+9 {\bf q}^{24} {\bf t}^{20}+{\bf q}^{26} {\bf t}^{18}+{\bf q}^{22} {\bf t}^{24}+19 {\bf q}^{24} {\bf t}^{22}+5 {\bf q}^{26} {\bf t}^{20}+10 {\bf q}^{24} {\bf t}^{24}+16 {\bf q}^{26} {\bf t}^{22}+2 {\bf q}^{28} {\bf t}^{20}+25 {\bf q}^{26} {\bf t}^{24}+9 {\bf q}^{28} {\bf t}^{22}+{\bf q}^{30} {\bf t}^{20}+8 {\bf q}^{26} {\bf t}^{26}+24 {\bf q}^{28} {\bf t}^{24}+5 {\bf q}^{30} {\bf t}^{22}+26 {\bf q}^{28} {\bf t}^{26}+17 {\bf q}^{30} {\bf t}^{24}+2 {\bf q}^{32} {\bf t}^{22}+5 {\bf q}^{28} {\bf t}^{28}+34 {\bf q}^{30} {\bf t}^{26}+9 {\bf q}^{32} {\bf t}^{24}+{\bf q}^{34} {\bf t}^{22}+26 {\bf q}^{30} {\bf t}^{28}+26 {\bf q}^{32} {\bf t}^{26}+5 {\bf q}^{34} {\bf t}^{24}+2 {\bf q}^{30} {\bf t}^{30}+42 {\bf q}^{32} {\bf t}^{28}+17 {\bf q}^{34} {\bf t}^{26}+2 {\bf q}^{36} {\bf t}^{24}+19 {\bf q}^{32} {\bf t}^{30}+39 {\bf q}^{34} {\bf t}^{28}+9 {\bf q}^{36} {\bf t}^{26}+{\bf q}^{38} {\bf t}^{24}+{\bf q}^{32} {\bf t}^{32}+47 {\bf q}^{34} {\bf t}^{30}+27 {\bf q}^{36} {\bf t}^{28}+5 {\bf q}^{38} {\bf t}^{26}+13 {\bf q}^{34} {\bf t}^{32}+51 {\bf q}^{36} {\bf t}^{30}+17 {\bf q}^{38} {\bf t}^{28}+2 {\bf q}^{40} {\bf t}^{26}+43 {\bf q}^{36} {\bf t}^{32}+41 {\bf q}^{38} {\bf t}^{30}+9 {\bf q}^{40} {\bf t}^{28}+{\bf q}^{42} {\bf t}^{26}+5 {\bf q}^{36} {\bf t}^{34}+64 {\bf q}^{38} {\bf t}^{32}+27 {\bf q}^{40} {\bf t}^{30}+5 {\bf q}^{42} {\bf t}^{28}+32 {\bf q}^{38} {\bf t}^{34}+56 {\bf q}^{40} {\bf t}^{32}+17 {\bf q}^{42} {\bf t}^{30}+2 {\bf q}^{44} {\bf t}^{28}+2 {\bf q}^{38} {\bf t}^{36}+65 {\bf q}^{40} {\bf t}^{34}+42 {\bf q}^{42} {\bf t}^{32}+9 {\bf q}^{44} {\bf t}^{30}+{\bf q}^{46} {\bf t}^{28}+18 {\bf q}^{40} {\bf t}^{36}+73 {\bf q}^{42} {\bf t}^{34}+27 {\bf q}^{44} {\bf t}^{32}+5 {\bf q}^{46} {\bf t}^{30}+56 {\bf q}^{42} {\bf t}^{36}+58 {\bf q}^{44} {\bf t}^{34}+17 {\bf q}^{46} {\bf t}^{32}+2 {\bf q}^{48} {\bf t}^{30}+7 {\bf q}^{42} {\bf t}^{38}+81 {\bf q}^{44} {\bf t}^{36}+42 {\bf q}^{46} {\bf t}^{34}+9 {\bf q}^{48} {\bf t}^{32}+{\bf q}^{50} {\bf t}^{30}+34 {\bf q}^{44} {\bf t}^{38}+78 {\bf q}^{46} {\bf t}^{36}+27 {\bf q}^{48} {\bf t}^{34}+5 {\bf q}^{50} {\bf t}^{32}+2 {\bf q}^{44} {\bf t}^{40}+75 {\bf q}^{46} {\bf t}^{38}+59 {\bf q}^{48} {\bf t}^{36}+17 {\bf q}^{50} {\bf t}^{34}+2 {\bf q}^{52} {\bf t}^{32}+16 {\bf q}^{46} {\bf t}^{40}+90 {\bf q}^{48} {\bf t}^{38}+42 {\bf q}^{50} {\bf t}^{36}+9 {\bf q}^{52} {\bf t}^{34}+{\bf q}^{54} {\bf t}^{32}+50 {\bf q}^{48} {\bf t}^{40}+80 {\bf q}^{50} {\bf t}^{38}+27 {\bf q}^{52} {\bf t}^{36}+5 {\bf q}^{54} {\bf t}^{34}+4 {\bf q}^{48} {\bf t}^{42}+87 {\bf q}^{50} {\bf t}^{40}+59 {\bf q}^{52} {\bf t}^{38}+17 {\bf q}^{54} {\bf t}^{36}+2 {\bf q}^{56} {\bf t}^{34}+23 {\bf q}^{50} {\bf t}^{42}+94 {\bf q}^{52} {\bf t}^{40}+42 {\bf q}^{54} {\bf t}^{38}+9 {\bf q}^{56} {\bf t}^{36}+{\bf q}^{58} {\bf t}^{34}+{\bf q}^{50} {\bf t}^{44}+58 {\bf q}^{52} {\bf t}^{42}+81 {\bf q}^{54} {\bf t}^{40}+27 {\bf q}^{56} {\bf t}^{38}+5 {\bf q}^{58} {\bf t}^{36}+7 {\bf q}^{52} {\bf t}^{44}+91 {\bf q}^{54} {\bf t}^{42}+59 {\bf q}^{56} {\bf t}^{40}+17 {\bf q}^{58} {\bf t}^{38}+2 {\bf q}^{60} {\bf t}^{36}+26 {\bf q}^{54} {\bf t}^{44}+94 {\bf q}^{56} {\bf t}^{42}+42 {\bf q}^{58} {\bf t}^{40}+9 {\bf q}^{60} {\bf t}^{38}+{\bf q}^{62} {\bf t}^{36}+{\bf q}^{54} {\bf t}^{46}+58 {\bf q}^{56} {\bf t}^{44}+80 {\bf q}^{58} {\bf t}^{42}+27 {\bf q}^{60} {\bf t}^{40}+5 {\bf q}^{62} {\bf t}^{38}+7 {\bf q}^{56} {\bf t}^{46}+87 {\bf q}^{58} {\bf t}^{44}+59 {\bf q}^{60} {\bf t}^{42}+17 {\bf q}^{62} {\bf t}^{40}+2 {\bf q}^{64} {\bf t}^{38}+23 {\bf q}^{58} {\bf t}^{46}+90 {\bf q}^{60} {\bf t}^{44}+42 {\bf q}^{62} {\bf t}^{42}+9 {\bf q}^{64} {\bf t}^{40}+{\bf q}^{66} {\bf t}^{38}+{\bf q}^{58} {\bf t}^{48}+50 {\bf q}^{60} {\bf t}^{46}+78 {\bf q}^{62} {\bf t}^{44}+27 {\bf q}^{64} {\bf t}^{42}+5 {\bf q}^{66} {\bf t}^{40}+4 {\bf q}^{60} {\bf t}^{48}+75 {\bf q}^{62} {\bf t}^{46}+58 {\bf q}^{64} {\bf t}^{44}+17 {\bf q}^{66} {\bf t}^{42}+2 {\bf q}^{68} {\bf t}^{40}+16 {\bf q}^{62} {\bf t}^{48}+81 {\bf q}^{64} {\bf t}^{46}+42 {\bf q}^{66} {\bf t}^{44}+9 {\bf q}^{68} {\bf t}^{42}+{\bf q}^{70} {\bf t}^{40}+34 {\bf q}^{64} {\bf t}^{48}+73 {\bf q}^{66} {\bf t}^{46}+27 {\bf q}^{68} {\bf t}^{44}+5 {\bf q}^{70} {\bf t}^{42}+2 {\bf q}^{64} {\bf t}^{50}+56 {\bf q}^{66} {\bf t}^{48}+56 {\bf q}^{68} {\bf t}^{46}+17 {\bf q}^{70} {\bf t}^{44}+2 {\bf q}^{72} {\bf t}^{42}+7 {\bf q}^{66} {\bf t}^{50}+65 {\bf q}^{68} {\bf t}^{48}+41 {\bf q}^{70} {\bf t}^{46}+9 {\bf q}^{72} {\bf t}^{44}+{\bf q}^{74} {\bf t}^{42}+18 {\bf q}^{68} {\bf t}^{50}+64 {\bf q}^{70} {\bf t}^{48}+27 {\bf q}^{72} {\bf t}^{46}+5 {\bf q}^{74} {\bf t}^{44}+32 {\bf q}^{70} {\bf t}^{50}+51 {\bf q}^{72} {\bf t}^{48}+17 {\bf q}^{74} {\bf t}^{46}+2 {\bf q}^{76} {\bf t}^{44}+2 {\bf q}^{70} {\bf t}^{52}+43 {\bf q}^{72} {\bf t}^{50}+39 {\bf q}^{74} {\bf t}^{48}+9 {\bf q}^{76} {\bf t}^{46}+{\bf q}^{78} {\bf t}^{44}+5 {\bf q}^{72} {\bf t}^{52}+47 {\bf q}^{74} {\bf t}^{50}+26 {\bf q}^{76} {\bf t}^{48}+5 {\bf q}^{78} {\bf t}^{46}+13 {\bf q}^{74} {\bf t}^{52}+42 {\bf q}^{76} {\bf t}^{50}+17 {\bf q}^{78} {\bf t}^{48}+2 {\bf q}^{80} {\bf t}^{46}+19 {\bf q}^{76} {\bf t}^{52}+34 {\bf q}^{78} {\bf t}^{50}+9 {\bf q}^{80} {\bf t}^{48}+{\bf q}^{82} {\bf t}^{46}+{\bf q}^{76} {\bf t}^{54}+26 {\bf q}^{78} {\bf t}^{52}+24 {\bf q}^{80} {\bf t}^{50}+5 {\bf q}^{82} {\bf t}^{48}+2 {\bf q}^{78} {\bf t}^{54}+26 {\bf q}^{80} {\bf t}^{52}+16 {\bf q}^{82} {\bf t}^{50}+2 {\bf q}^{84} {\bf t}^{48}+5 {\bf q}^{80} {\bf t}^{54}+25 {\bf q}^{82} {\bf t}^{52}+9 {\bf q}^{84} {\bf t}^{50}+{\bf q}^{86} {\bf t}^{48}+8 {\bf q}^{82} {\bf t}^{54}+19 {\bf q}^{84} {\bf t}^{52}+5 {\bf q}^{86} {\bf t}^{50}+10 {\bf q}^{84} {\bf t}^{54}+14 {\bf q}^{86} {\bf t}^{52}+2 {\bf q}^{88} {\bf t}^{50}+11 {\bf q}^{86} {\bf t}^{54}+8 {\bf q}^{88} {\bf t}^{52}+{\bf q}^{90} {\bf t}^{50}+{\bf q}^{86} {\bf t}^{56}+11 {\bf q}^{88} {\bf t}^{54}+5 {\bf q}^{90} {\bf t}^{52}+{\bf q}^{88} {\bf t}^{56}+9 {\bf q}^{90} {\bf t}^{54}+2 {\bf q}^{92} {\bf t}^{52}+2 {\bf q}^{90} {\bf t}^{56}+6 {\bf q}^{92} {\bf t}^{54}+{\bf q}^{94} {\bf t}^{52}+2 {\bf q}^{92} {\bf t}^{56}+4 {\bf q}^{94} {\bf t}^{54}+3 {\bf q}^{94} {\bf t}^{56}+2 {\bf q}^{96} {\bf t}^{54}+2 {\bf q}^{96} {\bf t}^{56}+{\bf q}^{98} {\bf t}^{54}+2 {\bf q}^{98} {\bf t}^{56}+{\bf q}^{100} {\bf t}^{56}+{\bf q}^{102} {\bf t}^{56}) {\bf a}^4+({\bf q}^{12} {\bf t}^{15}+{\bf q}^{14} {\bf t}^{17}+{\bf q}^{16} {\bf t}^{17}+2 {\bf q}^{16} {\bf t}^{19}+2 {\bf q}^{18} {\bf t}^{19}+3 {\bf q}^{18} {\bf t}^{21}+{\bf q}^{20} {\bf t}^{19}+4 {\bf q}^{20} {\bf t}^{21}+3 {\bf q}^{20} {\bf t}^{23}+2 {\bf q}^{22} {\bf t}^{21}+7 {\bf q}^{22} {\bf t}^{23}+{\bf q}^{24} {\bf t}^{21}+3 {\bf q}^{22} {\bf t}^{25}+5 {\bf q}^{24} {\bf t}^{23}+10 {\bf q}^{24} {\bf t}^{25}+2 {\bf q}^{26} {\bf t}^{23}+3 {\bf q}^{24} {\bf t}^{27}+9 {\bf q}^{26} {\bf t}^{25}+{\bf q}^{28} {\bf t}^{23}+12 {\bf q}^{26} {\bf t}^{27}+5 {\bf q}^{28} {\bf t}^{25}+2 {\bf q}^{26} {\bf t}^{29}+15 {\bf q}^{28} {\bf t}^{27}+2 {\bf q}^{30} {\bf t}^{25}+13 {\bf q}^{28} {\bf t}^{29}+10 {\bf q}^{30} {\bf t}^{27}+{\bf q}^{32} {\bf t}^{25}+{\bf q}^{28} {\bf t}^{31}+21 {\bf q}^{30} {\bf t}^{29}+5 {\bf q}^{32} {\bf t}^{27}+12 {\bf q}^{30} {\bf t}^{31}+17 {\bf q}^{32} {\bf t}^{29}+2 {\bf q}^{34} {\bf t}^{27}+{\bf q}^{30} {\bf t}^{33}+27 {\bf q}^{32} {\bf t}^{31}+10 {\bf q}^{34} {\bf t}^{29}+{\bf q}^{36} {\bf t}^{27}+9 {\bf q}^{32} {\bf t}^{33}+26 {\bf q}^{34} {\bf t}^{31}+5 {\bf q}^{36} {\bf t}^{29}+28 {\bf q}^{34} {\bf t}^{33}+18 {\bf q}^{36} {\bf t}^{31}+2 {\bf q}^{38} {\bf t}^{29}+6 {\bf q}^{34} {\bf t}^{35}+37 {\bf q}^{36} {\bf t}^{33}+10 {\bf q}^{38} {\bf t}^{31}+{\bf q}^{40} {\bf t}^{29}+26 {\bf q}^{36} {\bf t}^{35}+28 {\bf q}^{38} {\bf t}^{33}+5 {\bf q}^{40} {\bf t}^{31}+3 {\bf q}^{36} {\bf t}^{37}+44 {\bf q}^{38} {\bf t}^{35}+18 {\bf q}^{40} {\bf t}^{33}+2 {\bf q}^{42} {\bf t}^{31}+19 {\bf q}^{38} {\bf t}^{37}+42 {\bf q}^{40} {\bf t}^{35}+10 {\bf q}^{42} {\bf t}^{33}+{\bf q}^{44} {\bf t}^{31}+{\bf q}^{38} {\bf t}^{39}+45 {\bf q}^{40} {\bf t}^{37}+29 {\bf q}^{42} {\bf t}^{35}+5 {\bf q}^{44} {\bf t}^{33}+11 {\bf q}^{40} {\bf t}^{39}+54 {\bf q}^{42} {\bf t}^{37}+18 {\bf q}^{44} {\bf t}^{35}+2 {\bf q}^{46} {\bf t}^{33}+38 {\bf q}^{42} {\bf t}^{39}+44 {\bf q}^{44} {\bf t}^{37}+10 {\bf q}^{46} {\bf t}^{35}+{\bf q}^{48} {\bf t}^{33}+5 {\bf q}^{42} {\bf t}^{41}+61 {\bf q}^{44} {\bf t}^{39}+29 {\bf q}^{46} {\bf t}^{37}+5 {\bf q}^{48} {\bf t}^{35}+24 {\bf q}^{44} {\bf t}^{41}+59 {\bf q}^{46} {\bf t}^{39}+18 {\bf q}^{48} {\bf t}^{37}+2 {\bf q}^{50} {\bf t}^{35}+{\bf q}^{44} {\bf t}^{43}+54 {\bf q}^{46} {\bf t}^{41}+45 {\bf q}^{48} {\bf t}^{39}+10 {\bf q}^{50} {\bf t}^{37}+{\bf q}^{52} {\bf t}^{35}+12 {\bf q}^{46} {\bf t}^{43}+70 {\bf q}^{48} {\bf t}^{41}+29 {\bf q}^{50} {\bf t}^{39}+5 {\bf q}^{52} {\bf t}^{37}+37 {\bf q}^{48} {\bf t}^{43}+61 {\bf q}^{50} {\bf t}^{41}+18 {\bf q}^{52} {\bf t}^{39}+2 {\bf q}^{54} {\bf t}^{37}+4 {\bf q}^{48} {\bf t}^{45}+66 {\bf q}^{50} {\bf t}^{43}+45 {\bf q}^{52} {\bf t}^{41}+10 {\bf q}^{54} {\bf t}^{39}+{\bf q}^{56} {\bf t}^{37}+17 {\bf q}^{50} {\bf t}^{45}+74 {\bf q}^{52} {\bf t}^{43}+29 {\bf q}^{54} {\bf t}^{41}+5 {\bf q}^{56} {\bf t}^{39}+{\bf q}^{50} {\bf t}^{47}+44 {\bf q}^{52} {\bf t}^{45}+62 {\bf q}^{54} {\bf t}^{43}+18 {\bf q}^{56} {\bf t}^{41}+2 {\bf q}^{58} {\bf t}^{39}+6 {\bf q}^{52} {\bf t}^{47}+69 {\bf q}^{54} {\bf t}^{45}+45 {\bf q}^{56} {\bf t}^{43}+10 {\bf q}^{58} {\bf t}^{41}+{\bf q}^{60} {\bf t}^{39}+21 {\bf q}^{54} {\bf t}^{47}+74 {\bf q}^{56} {\bf t}^{45}+29 {\bf q}^{58} {\bf t}^{43}+5 {\bf q}^{60} {\bf t}^{41}+{\bf q}^{54} {\bf t}^{49}+44 {\bf q}^{56} {\bf t}^{47}+61 {\bf q}^{58} {\bf t}^{45}+18 {\bf q}^{60} {\bf t}^{43}+2 {\bf q}^{62} {\bf t}^{41}+6 {\bf q}^{56} {\bf t}^{49}+66 {\bf q}^{58} {\bf t}^{47}+45 {\bf q}^{60} {\bf t}^{45}+10 {\bf q}^{62} {\bf t}^{43}+{\bf q}^{64} {\bf t}^{41}+17 {\bf q}^{58} {\bf t}^{49}+70 {\bf q}^{60} {\bf t}^{47}+29 {\bf q}^{62} {\bf t}^{45}+5 {\bf q}^{64} {\bf t}^{43}+{\bf q}^{58} {\bf t}^{51}+37 {\bf q}^{60} {\bf t}^{49}+59 {\bf q}^{62} {\bf t}^{47}+18 {\bf q}^{64} {\bf t}^{45}+2 {\bf q}^{66} {\bf t}^{43}+4 {\bf q}^{60} {\bf t}^{51}+54 {\bf q}^{62} {\bf t}^{49}+44 {\bf q}^{64} {\bf t}^{47}+10 {\bf q}^{66} {\bf t}^{45}+{\bf q}^{68} {\bf t}^{43}+12 {\bf q}^{62} {\bf t}^{51}+61 {\bf q}^{64} {\bf t}^{49}+29 {\bf q}^{66} {\bf t}^{47}+5 {\bf q}^{68} {\bf t}^{45}+24 {\bf q}^{64} {\bf t}^{51}+54 {\bf q}^{66} {\bf t}^{49}+18 {\bf q}^{68} {\bf t}^{47}+2 {\bf q}^{70} {\bf t}^{45}+{\bf q}^{64} {\bf t}^{53}+38 {\bf q}^{66} {\bf t}^{51}+42 {\bf q}^{68} {\bf t}^{49}+10 {\bf q}^{70} {\bf t}^{47}+{\bf q}^{72} {\bf t}^{45}+5 {\bf q}^{66} {\bf t}^{53}+45 {\bf q}^{68} {\bf t}^{51}+28 {\bf q}^{70} {\bf t}^{49}+5 {\bf q}^{72} {\bf t}^{47}+11 {\bf q}^{68} {\bf t}^{53}+44 {\bf q}^{70} {\bf t}^{51}+18 {\bf q}^{72} {\bf t}^{49}+2 {\bf q}^{74} {\bf t}^{47}+19 {\bf q}^{70} {\bf t}^{53}+37 {\bf q}^{72} {\bf t}^{51}+10 {\bf q}^{74} {\bf t}^{49}+{\bf q}^{76} {\bf t}^{47}+{\bf q}^{70} {\bf t}^{55}+26 {\bf q}^{72} {\bf t}^{53}+26 {\bf q}^{74} {\bf t}^{51}+5 {\bf q}^{76} {\bf t}^{49}+3 {\bf q}^{72} {\bf t}^{55}+28 {\bf q}^{74} {\bf t}^{53}+17 {\bf q}^{76} {\bf t}^{51}+2 {\bf q}^{78} {\bf t}^{49}+6 {\bf q}^{74} {\bf t}^{55}+27 {\bf q}^{76} {\bf t}^{53}+10 {\bf q}^{78} {\bf t}^{51}+{\bf q}^{80} {\bf t}^{49}+9 {\bf q}^{76} {\bf t}^{55}+21 {\bf q}^{78} {\bf t}^{53}+5 {\bf q}^{80} {\bf t}^{51}+12 {\bf q}^{78} {\bf t}^{55}+15 {\bf q}^{80} {\bf t}^{53}+2 {\bf q}^{82} {\bf t}^{51}+{\bf q}^{78} {\bf t}^{57}+13 {\bf q}^{80} {\bf t}^{55}+9 {\bf q}^{82} {\bf t}^{53}+{\bf q}^{84} {\bf t}^{51}+{\bf q}^{80} {\bf t}^{57}+12 {\bf q}^{82} {\bf t}^{55}+5 {\bf q}^{84} {\bf t}^{53}+2 {\bf q}^{82} {\bf t}^{57}+10 {\bf q}^{84} {\bf t}^{55}+2 {\bf q}^{86} {\bf t}^{53}+3 {\bf q}^{84} {\bf t}^{57}+7 {\bf q}^{86} {\bf t}^{55}+{\bf q}^{88} {\bf t}^{53}+3 {\bf q}^{86} {\bf t}^{57}+4 {\bf q}^{88} {\bf t}^{55}+3 {\bf q}^{88} {\bf t}^{57}+2 {\bf q}^{90} {\bf t}^{55}+3 {\bf q}^{90} {\bf t}^{57}+{\bf q}^{92} {\bf t}^{55}+2 {\bf q}^{92} {\bf t}^{57}+{\bf q}^{94} {\bf t}^{57}+{\bf q}^{96} {\bf t}^{57}) {\bf a}^6+({\bf q}^{20} {\bf t}^{24}+{\bf q}^{22} {\bf t}^{26}+{\bf q}^{24} {\bf t}^{26}+2 {\bf q}^{24} {\bf t}^{28}+2 {\bf q}^{26} {\bf t}^{28}+2 {\bf q}^{26} {\bf t}^{30}+{\bf q}^{28} {\bf t}^{28}+4 {\bf q}^{28} {\bf t}^{30}+3 {\bf q}^{28} {\bf t}^{32}+2 {\bf q}^{30} {\bf t}^{30}+6 {\bf q}^{30} {\bf t}^{32}+{\bf q}^{32} {\bf t}^{30}+2 {\bf q}^{30} {\bf t}^{34}+5 {\bf q}^{32} {\bf t}^{32}+8 {\bf q}^{32} {\bf t}^{34}+2 {\bf q}^{34} {\bf t}^{32}+2 {\bf q}^{32} {\bf t}^{36}+8 {\bf q}^{34} {\bf t}^{34}+{\bf q}^{36} {\bf t}^{32}+9 {\bf q}^{34} {\bf t}^{36}+5 {\bf q}^{36} {\bf t}^{34}+{\bf q}^{34} {\bf t}^{38}+13 {\bf q}^{36} {\bf t}^{36}+2 {\bf q}^{38} {\bf t}^{34}+8 {\bf q}^{36} {\bf t}^{38}+9 {\bf q}^{38} {\bf t}^{36}+{\bf q}^{40} {\bf t}^{34}+{\bf q}^{36} {\bf t}^{40}+16 {\bf q}^{38} {\bf t}^{38}+5 {\bf q}^{40} {\bf t}^{36}+6 {\bf q}^{38} {\bf t}^{40}+15 {\bf q}^{40} {\bf t}^{38}+2 {\bf q}^{42} {\bf t}^{36}+18 {\bf q}^{40} {\bf t}^{40}+9 {\bf q}^{42} {\bf t}^{38}+{\bf q}^{44} {\bf t}^{36}+4 {\bf q}^{40} {\bf t}^{42}+21 {\bf q}^{42} {\bf t}^{40}+5 {\bf q}^{44} {\bf t}^{38}+14 {\bf q}^{42} {\bf t}^{42}+16 {\bf q}^{44} {\bf t}^{40}+2 {\bf q}^{46} {\bf t}^{38}+2 {\bf q}^{42} {\bf t}^{44}+25 {\bf q}^{44} {\bf t}^{42}+9 {\bf q}^{46} {\bf t}^{40}+{\bf q}^{48} {\bf t}^{38}+11 {\bf q}^{44} {\bf t}^{44}+23 {\bf q}^{46} {\bf t}^{42}+5 {\bf q}^{48} {\bf t}^{40}+23 {\bf q}^{46} {\bf t}^{44}+16 {\bf q}^{48} {\bf t}^{42}+2 {\bf q}^{50} {\bf t}^{40}+5 {\bf q}^{46} {\bf t}^{46}+30 {\bf q}^{48} {\bf t}^{44}+9 {\bf q}^{50} {\bf t}^{42}+{\bf q}^{52} {\bf t}^{40}+17 {\bf q}^{48} {\bf t}^{46}+24 {\bf q}^{50} {\bf t}^{44}+5 {\bf q}^{52} {\bf t}^{42}+2 {\bf q}^{48} {\bf t}^{48}+28 {\bf q}^{50} {\bf t}^{46}+16 {\bf q}^{52} {\bf t}^{44}+2 {\bf q}^{54} {\bf t}^{42}+9 {\bf q}^{50} {\bf t}^{48}+32 {\bf q}^{52} {\bf t}^{46}+9 {\bf q}^{54} {\bf t}^{44}+{\bf q}^{56} {\bf t}^{42}+21 {\bf q}^{52} {\bf t}^{48}+24 {\bf q}^{54} {\bf t}^{46}+5 {\bf q}^{56} {\bf t}^{44}+3 {\bf q}^{52} {\bf t}^{50}+30 {\bf q}^{54} {\bf t}^{48}+16 {\bf q}^{56} {\bf t}^{46}+2 {\bf q}^{58} {\bf t}^{44}+10 {\bf q}^{54} {\bf t}^{50}+32 {\bf q}^{56} {\bf t}^{48}+9 {\bf q}^{58} {\bf t}^{46}+{\bf q}^{60} {\bf t}^{44}+{\bf q}^{54} {\bf t}^{52}+21 {\bf q}^{56} {\bf t}^{50}+24 {\bf q}^{58} {\bf t}^{48}+5 {\bf q}^{60} {\bf t}^{46}+3 {\bf q}^{56} {\bf t}^{52}+28 {\bf q}^{58} {\bf t}^{50}+16 {\bf q}^{60} {\bf t}^{48}+2 {\bf q}^{62} {\bf t}^{46}+9 {\bf q}^{58} {\bf t}^{52}+30 {\bf q}^{60} {\bf t}^{50}+9 {\bf q}^{62} {\bf t}^{48}+{\bf q}^{64} {\bf t}^{46}+17 {\bf q}^{60} {\bf t}^{52}+23 {\bf q}^{62} {\bf t}^{50}+5 {\bf q}^{64} {\bf t}^{48}+2 {\bf q}^{60} {\bf t}^{54}+23 {\bf q}^{62} {\bf t}^{52}+16 {\bf q}^{64} {\bf t}^{50}+2 {\bf q}^{66} {\bf t}^{48}+5 {\bf q}^{62} {\bf t}^{54}+25 {\bf q}^{64} {\bf t}^{52}+9 {\bf q}^{66} {\bf t}^{50}+{\bf q}^{68} {\bf t}^{48}+11 {\bf q}^{64} {\bf t}^{54}+21 {\bf q}^{66} {\bf t}^{52}+5 {\bf q}^{68} {\bf t}^{50}+14 {\bf q}^{66} {\bf t}^{54}+15 {\bf q}^{68} {\bf t}^{52}+2 {\bf q}^{70} {\bf t}^{50}+2 {\bf q}^{66} {\bf t}^{56}+18 {\bf q}^{68} {\bf t}^{54}+9 {\bf q}^{70} {\bf t}^{52}+{\bf q}^{72} {\bf t}^{50}+4 {\bf q}^{68} {\bf t}^{56}+16 {\bf q}^{70} {\bf t}^{54}+5 {\bf q}^{72} {\bf t}^{52}+6 {\bf q}^{70} {\bf t}^{56}+13 {\bf q}^{72} {\bf t}^{54}+2 {\bf q}^{74} {\bf t}^{52}+8 {\bf q}^{72} {\bf t}^{56}+8 {\bf q}^{74} {\bf t}^{54}+{\bf q}^{76} {\bf t}^{52}+{\bf q}^{72} {\bf t}^{58}+9 {\bf q}^{74} {\bf t}^{56}+5 {\bf q}^{76} {\bf t}^{54}+{\bf q}^{74} {\bf t}^{58}+8 {\bf q}^{76} {\bf t}^{56}+2 {\bf q}^{78} {\bf t}^{54}+2 {\bf q}^{76} {\bf t}^{58}+6 {\bf q}^{78} {\bf t}^{56}+{\bf q}^{80} {\bf t}^{54}+2 {\bf q}^{78} {\bf t}^{58}+4 {\bf q}^{80} {\bf t}^{56}+3 {\bf q}^{80} {\bf t}^{58}+2 {\bf q}^{82} {\bf t}^{56}+2 {\bf q}^{82} {\bf t}^{58}+{\bf q}^{84} {\bf t}^{56}+2 {\bf q}^{84} {\bf t}^{58}+{\bf q}^{86} {\bf t}^{58}+{\bf q}^{88} {\bf t}^{58}) {\bf a}^8+({\bf q}^{30} {\bf t}^{35}+{\bf q}^{32} {\bf t}^{37}+{\bf q}^{34} {\bf t}^{37}+{\bf q}^{34} {\bf t}^{39}+2 {\bf q}^{36} {\bf t}^{39}+{\bf q}^{36} {\bf t}^{41}+{\bf q}^{38} {\bf t}^{39}+3 {\bf q}^{38} {\bf t}^{41}+{\bf q}^{38} {\bf t}^{43}+2 {\bf q}^{40} {\bf t}^{41}+3 {\bf q}^{40} {\bf t}^{43}+{\bf q}^{42} {\bf t}^{41}+{\bf q}^{40} {\bf t}^{45}+4 {\bf q}^{42} {\bf t}^{43}+3 {\bf q}^{42} {\bf t}^{45}+2 {\bf q}^{44} {\bf t}^{43}+5 {\bf q}^{44} {\bf t}^{45}+{\bf q}^{46} {\bf t}^{43}+3 {\bf q}^{44} {\bf t}^{47}+4 {\bf q}^{46} {\bf t}^{45}+6 {\bf q}^{46} {\bf t}^{47}+2 {\bf q}^{48} {\bf t}^{45}+{\bf q}^{46} {\bf t}^{49}+6 {\bf q}^{48} {\bf t}^{47}+{\bf q}^{50} {\bf t}^{45}+5 {\bf q}^{48} {\bf t}^{49}+4 {\bf q}^{50} {\bf t}^{47}+7 {\bf q}^{50} {\bf t}^{49}+2 {\bf q}^{52} {\bf t}^{47}+3 {\bf q}^{50} {\bf t}^{51}+6 {\bf q}^{52} {\bf t}^{49}+{\bf q}^{54} {\bf t}^{47}+6 {\bf q}^{52} {\bf t}^{51}+4 {\bf q}^{54} {\bf t}^{49}+{\bf q}^{52} {\bf t}^{53}+8 {\bf q}^{54} {\bf t}^{51}+2 {\bf q}^{56} {\bf t}^{49}+3 {\bf q}^{54} {\bf t}^{53}+6 {\bf q}^{56} {\bf t}^{51}+{\bf q}^{58} {\bf t}^{49}+6 {\bf q}^{56} {\bf t}^{53}+4 {\bf q}^{58} {\bf t}^{51}+{\bf q}^{56} {\bf t}^{55}+7 {\bf q}^{58} {\bf t}^{53}+2 {\bf q}^{60} {\bf t}^{51}+3 {\bf q}^{58} {\bf t}^{55}+6 {\bf q}^{60} {\bf t}^{53}+{\bf q}^{62} {\bf t}^{51}+5 {\bf q}^{60} {\bf t}^{55}+4 {\bf q}^{62} {\bf t}^{53}+6 {\bf q}^{62} {\bf t}^{55}+2 {\bf q}^{64} {\bf t}^{53}+{\bf q}^{62} {\bf t}^{57}+5 {\bf q}^{64} {\bf t}^{55}+{\bf q}^{66} {\bf t}^{53}+3 {\bf q}^{64} {\bf t}^{57}+4 {\bf q}^{66} {\bf t}^{55}+3 {\bf q}^{66} {\bf t}^{57}+2 {\bf q}^{68} {\bf t}^{55}+3 {\bf q}^{68} {\bf t}^{57}+{\bf q}^{70} {\bf t}^{55}+{\bf q}^{68} {\bf t}^{59}+3 {\bf q}^{70} {\bf t}^{57}+{\bf q}^{70} {\bf t}^{59}+2 {\bf q}^{72} {\bf t}^{57}+{\bf q}^{72} {\bf t}^{59}+{\bf q}^{74} {\bf t}^{57}+{\bf q}^{74} {\bf t}^{59}+{\bf q}^{76} {\bf t}^{59}+{\bf q}^{78} {\bf t}^{59}) {\bf a}^{10}+({\bf q}^{42} {\bf t}^{48}+{\bf q}^{46} {\bf t}^{50}+{\bf q}^{48} {\bf t}^{52}+{\bf q}^{50} {\bf t}^{52}+{\bf q}^{52} {\bf t}^{54}+{\bf q}^{54} {\bf t}^{54}+{\bf q}^{54} {\bf t}^{56}+{\bf q}^{56} {\bf t}^{56}+{\bf q}^{58} {\bf t}^{56}+{\bf q}^{60} {\bf t}^{58}+{\bf q}^{62} {\bf t}^{58}+{\bf q}^{66} {\bf t}^{60}) {\bf a}^{12}
$

\listrefs
\end